\documentclass[prd,aps,floats,preprint]{revtex4}
\usepackage{longtable}
\usepackage{graphicx}
\usepackage{amsmath,amssymb}
\usepackage{color}
\usepackage{bbm}

\newcommand{\braket}[2]{\left\langle#1\, |\,#2\,\right\rangle}  %  < #1 | #2 >
\newcommand{\expec}[1]{\langle#1\rangle}  %  < #1 >
\newcommand{\e}{{\rm e}}
\newcommand{\drm}{{\rm d}}
\newcommand{\irm}{{\rm i}}
\newcommand{\beq}{\begin{equation}}
\newcommand{\eeq}{\end{equation}}
\newcommand{\bdm}{\begin{displaymath}}
\newcommand{\edm}{\end{displaymath}}
\newcommand{\T}[1]{\tilde{#1}}

\newcommand{\SNR}{\textnormal{SNR}}

\graphicspath{{./pics/}}

\begin{document}

\title{Subtraction-noise projection in gravitational-wave detector networks}

\author{Jan Harms}
\author{Christoph Mahrdt}
\author{Markus Otto}
\author{Malte Prie\ss}
\affiliation{Institut f\"ur Gravitationsphysik, Universit\"at Hannover and Max-Planck-Institut f\"ur Gravitationsphysik (Albert-Einstein-Institut), Callinstra\ss e~38, 30167 Hannover, Germany}

%% for REVTeX4, each author name can be set in a separate \author{} field

\date{\today}

\begin{abstract}
In this paper, we present a successful implementation of a subtraction-noise projection method into a simple, simulated data analysis pipeline of a gravitational-wave search. We investigate the problem to reveal a weak stochastic background signal which is covered by a strong foreground of compact-binary coalescences. The foreground which is estimated by matched filters, has to be subtracted from the data. Even an optimal analysis of foreground signals will leave subtraction noise due to estimation errors of template parameters which may corrupt the measurement of the background signal. The subtraction noise can be removed by a noise projection. We apply our analysis pipeline to the proposed future-generation space-borne Big Bang Observer (BBO) mission which seeks for a stochastic background of primordial GWs in the frequency range $\sim 0.1\,$Hz --- $1\,$Hz covered by a foreground of black-hole and neutron-star binaries. Our analysis is based on a simulation code which provides a dynamical model of a time-delay interferometer (TDI) network. It generates the data as time series and incorporates the analysis pipeline together with the noise projection. Our results confirm previous ad hoc predictions which say that BBO will be sensitive to backgrounds with fractional energy densities below $\Omega=10^{-16}$.
\end{abstract}

\maketitle %% NULL FUNCTION WITH LATEX 2e

\section{Introduction}
Currently, the first generation of large-scale laser interferometers is being operated to make a direct detection of a gravitational wave (GW) \cite{WiEA2002,LSC2007c,AcEA2007}. The primary targets of these detectors are compact-binary coalescences (CBCs), pulsars and supernovae. However, the predicted event rate is so small that a detection of GWs by these instruments is highly unlikely. Their science goals are to push technological developments for a next generation of detectors and to place upper limits on GW amplitudes, thereby deriving at least to some degree restrictions on astrophysical processes. These limits either refer to deterministic sources like pulsars and binaries \cite{LSC2005,LSC2007d} or to stochastic backgrounds of GWs which may have an astrophysical or cosmological origin \cite{LSC2007a}. Although limits on stochastic backgrounds already become (weakly) scientifically relevant, the current upper limit of the background energy is about 10 orders of magnitude above a likely value for the cosmological background assuming standard inflationary models. At this stage, confidence in a detection event would be significantly increased by combining the data of many different detectors \cite{BeEA2006b}. This technique has become a standard tool in GW data analysis, and it is the only method to coherently detect stochastic signals. Within ten years, the first generation of ground-based detectors will be joined or replaced by advanced LIGO  -- a second-generation ground-based detector -- and LISA, which will be the first space-borne laser-interferometric GW detector \cite{GuEA1999,LISA2000}. In contrast to first-generation detectors and advanced LIGO, LISA has to cope with a totally different data analysis problem. LISA will be sensitive to many sources which combine to form a GW signal foreground \cite{CrCo2004,RaCu2007}. This foreground is formed by millions of galactic white-dwarf (WD) binaries and cannot be resolved completely. The unresolved, residual foreground acts as Gaussian confusion noise which impairs the detection and analysis of other signals. Any future detector will have to take the source-confusion problem into account and find a way to solve or circumvent it.

Even if a signal foreground can be resolved, the estimated signal waveforms will deviate somewhat from the true signals due to instrumental or confusion noise. If the estimated waveforms are subtracted from the data, then a residual signal spectrum, the subtraction noise, remains. Recently, a method was proposed to remove the residual foreground under certain conditions \cite{CuHa2006}. This method is based on a geometrical interpretation of signal analysis. It allows to access a weak target like a stochastic GW background, irrespective of the fact that the residual foreground -- in that case resulting from inaccurate fitting of waveforms from binary neutron stars (NS) and black holes (BH) -- may be much stronger. The conditions which have to be fulfilled are that (1) an accurate model exists for the waveform of individual foreground signals, (2) the overlap between foreground and background signals is negligible or irrelevant, (3) there are not too many foreground sources compared to the amount of data being collected (this will be specified in section \ref{secProj}) and (4) the data is taken with a network consisting of at least 2 detectors. If the second condition is not fulfilled, then the removal of subtraction noise may deteriorate the waveforms of background signals in the data. The noise-removal algorithm which is geometrically defined as a noise projection, comes with many numerical challenges which could not be addressed in \cite{CuHa2006}. The purpose of this paper is to present a detailed discussion of the noise projection and to show how it can be implemented into a data analysis pipeline of a simulated future-generation detector network.

The network model of our simulation is based on a design draft for a future mission, the Big Bang Observer (BBO) \cite{Phi2003}. Its primary target is to measure the stochastic GW background with cosmological origin (CGWB) which was generated shortly after the big bang presumably during the inflationary phase \cite{All1996,SKC2006,BoBu2007}. For non-exotic (likely) models of the CGWB, the detector should be designed with peak sensitivity at lowest possible frequencies, since sensitivity towards stochastic backgrounds increases with decreasing signal frequency. This background will be overlayed at all frequencies by a foreground of CBCs which needs to be subtracted. At this point, one has to take into account the confusion noise problem. The galactic WD/WD foreground poses an intractable barrier even for future detectors. By consequence, its spectrum which reaches out to 0.25\,Hz \cite{FaPh2003}, sets a lower boundary on BBO's detection band. Lower frequencies beyond the WD/WD barrier are excluded by a foreground of merging supermassiv black holes and too less data would be collected at these frequencies. Above 0.25\,Hz, a remaining foreground of $10^5$ - $10^6$ NS and BH binaries has to be subtracted from the data. For BBO, the NS/NS mergers are the weakest foreground signals, and they are the most difficult to analyze and to subtract. Estimates for the number of merging NS/NS are highly uncertain. Extrapolating predictions of the galactic merger rate to the whole observable universe, one obtains values around $10^5$ for the NS/NS mergers per year \cite{SFMPZ2001,KNST2001,BKB2002}. As was explicitly shown in \cite{CuHa2006}, BBO is sensitive to virtually all NS and BH CBCs in the entire observable universe! Not surprisingly, detection and analysis of CBCs build the secondary target of BBO.

At an early stage of creating the simulation, it became clear that it would not be possible to demonstrate the projection method on a realistic foreground with $10^5$ or more events even if the search for the signals was excluded from the pipeline. Therefore, we chose to test the algorithm on a much smaller foreground consisting of 100 injected NS/NS systems. And even then it was necessary to shorten the observation time from the mission lifetime of 3\,yrs down to $10^5$\,s. In the end, our results have to be extrapolated to the full observation time of $10^8\,$s in order to derive a prediction of BBO's sensitivity towards the CGWB.

Our paper is organized as follows. In section \ref{secNetwork}, we describe in detail the fully-dynamical model of the detector network which underlies the simulation. In section \ref{secSimulate}, we give an overview of the simulation pipeline and highlight that the network design of BBO is tightly linked to the demands of the data analysis pipeline. A brief description of the signal model which determines the CBC waveforms and an introduction to the Fisher matrix which is one of the basic quantities of the projection method are given in section \ref{secFisher}. In section \ref{secStochastic}, we present a general framework how to simulate a stochastic signal in a network of space-borne detectors. The geometrical interpretation of statistics is outlined in section \ref{secProj} including a description of the subtraction-noise projection. The optimal cross-correlation scheme for BBO is explained and investigated in section \ref{secCross}. Results are given in section \ref{secResults} together with an extrapolation of BBO's sensitivity to an observation time which is equal to BBO's lifetime.

\section{The network model}
\label{secNetwork}
\begin{figure}[ht]
\includegraphics[width=8cm]{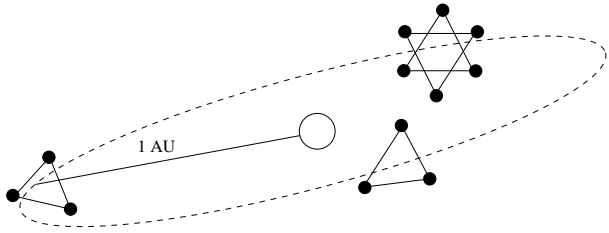}
\caption{The BBO network of LISA-type detectors}
\label{figBBOnetwork}
\end{figure}

\begin{figure}[ht]
\includegraphics[width=8cm]{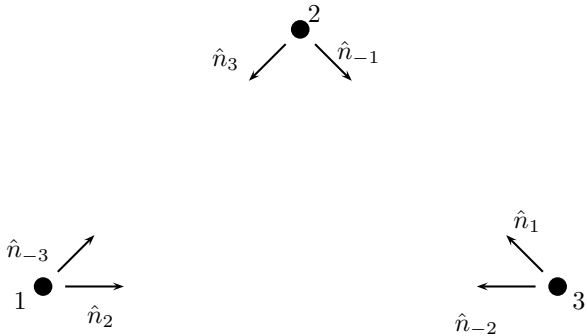}
\caption{Triangular BBO detector configuration}
\label{figBBOconf}
\end{figure}

BBO consists of four independent detectors which orbit the Sun at 1\,AU. Each detector is formed by three spacecrafts in a nearly equilateral triangular configuration (Fig.~\ref{figBBOnetwork}). The nominal distance between spacecrafts is $50000\,$km which entails that they follow slightly eccentric orbits with $e\sim 9.65\cdot 10^{-5}$. Each detector performs a cartwheel motion on the orbital path completing one rotation in one year. All triangles are tilted against the orbital plane by $60^\circ$. In addition, the relative distances between spacecrafts change by small amounts (0.01\%-0.02\%) during one year -- the so-called breathing motion -- and therefore the detectors cannot be treated as rigid objects. The motion of each detector can be described in a power series of the orbital eccentricity $e$ \cite{CoRu2003}. Expanding the exact orbital equations (which can be found in \cite{CoLi2007}) up to second order, the position vectors read
\begin{widetext}
\beq
\begin{split}
\vec r_{ij}(t)=&{\rm AU}\cdot 
\begin{pmatrix}
\cos(\alpha_i(t))\\[0.1cm]
\sin(\alpha_i(t))\\[0.1cm]
0
\end{pmatrix}\\[0.2cm]
&+e\cdot{\rm AU}\cdot
\begin{pmatrix}
\frac{1}{2}\cos(2\alpha_i(t)-\beta_{ij})-\frac{3}{2}\cos(\beta_{ij})\\[0.1cm]
\frac{1}{2}\sin(2\alpha_i(t)-\beta_{ij})-\frac{3}{2}\sin(\beta_{ij})\\[0.1cm]
-\sqrt{3}\cos(\alpha_i(t)-\beta_{ij})
\end{pmatrix}\\[0.2cm]
&+e^2\cdot{\rm AU}\cdot
\begin{pmatrix}
\frac{3}{8}\cos(3\alpha_i(t)-2\beta_{ij})-\frac{5}{8}\cos(\alpha_i(t)-2\beta_{ij})-\frac{5}{4}\cos(\alpha_i(t))\\[0.1cm]
\frac{3}{8}\sin(3\alpha_i(t)-2\beta_{ij})+\frac{5}{8}\sin(\alpha_i(t)-2\beta_{ij})-\frac{5}{4}\sin(\alpha_i(t))\\[0.1cm]
-\frac{1}{2}\sqrt{3}\left[\cos(2\alpha_i(t)-2\beta_{ij})-3\right]
\end{pmatrix}
\end{split}
\eeq
\end{widetext}
where the first-order correction of the orbital path corresponds to the detector's cartwheel motion and the second-order correction describes the small relative motion of the spacecrafts.  The angle $\alpha_i(t)=2\pi f_{\rm orb}t+\alpha_i(0)$ with $f_{\rm orb}=1/{\rm yr}$ determines the location of the detectors $i=1,\ldots,4$ on the orbit of the Earth and $\beta_{ij}=2(j-1)\pi/3+\xi_i$ fixes the position of spacecrafts $j=1,2,3$ in each detector $i$. The constants $\xi_i$ govern the relative cartwheel phases of detectors. In the following we will identify a spacecraft $ij$ by a single index $j$ assuming that all formulas are independently valid for each detector. The initial detector positions are $\vec\alpha(0)=(0,0,2\pi/3,4\pi/3)$ and the internal configuration for each detector is given by $\vec\xi=(0,\pi,0,0)$. A $\pi$ difference of the first two components of $\vec\xi$ puts the first two collocated detectors in a star-of-David configuration. Otherwise, components can be chosen arbitrarily.
A BBO spacecraft will certainly have a different design than a LISA spacecraft. However, it should be clear that the optimal sensitivity which is quantum-noise limited at high frequencies and acceleration-noise limited at low frequencies, does not depend on the topologies of the optical benches. Therefore, we assume a LISA-type optical-bench design of the BBO spacecrafts and make use of well-known LISA results to evaluate BBO's instrumental noise and GW response. For LISA, a minimum number of four photocurrents per spacecraft has to be included in a detector simulation. Two of them measure frequency fluctuations $y_{l}$ of the light coming from a neighboring spacecraft via detector arm $l$ and the other two measure intra-spacecraft signals which are denoted by $z_{l}$ where the photodiode (which records) $z_{l}$ is found on the same optical bench than the diode $y_{l}$ (there are two optical benches, one for each link to a neighboring spacecraft). The inter-spacecraft signals $y_l$ are sensitive to GWs. The link index $l$ assumes positive and negative values to discriminate between the two light-travelling directions $\hat n_l$ of the detector arm (see Fig.~\ref{figBBOconf}). 
Now, the noise spectrum of each photocurrent will be dominated by laser-frequency fluctuations and optical-bench noise \cite{TEA2002a}. The situation is different for ground-based detectors where laser noise interferes destructively at a beam splitter towards the output port and suspension systems and isolation schemes attenuate the equivalent of optical-bench noise. The solution is to establish destructive interference electronically by appropriately adding and subtracting photocurrents in each detector. Some photocurrents have to be added to others with certain time delays
\beq
\begin{split}
D_dy_l&\equiv y_{l,d}(t)\equiv y_l\left(t-L_d(t)/c\right)\\
D_{d_2}D_{d_1}y_l&\equiv y_{l;d_1d_2}(t)\\
&\equiv y_l\left(t-L_{d_2}(t)/c-L_{d_1}\left(t-L_{d_2}(t)/c\right)/c\right)
\end{split}
\label{eqDelay}
\eeq
where $L_d(t)$ is the length of the optical path of link $d$ which was travelled by light being detected at time $t$, and $c$ is the speed of light. That is why the electronic interference scheme is known as time-delay interferometry (TDI). We mention that one has to take into account that the light propagation directions $\hat n_l(t),\,-\hat n_{-l}(t)$ differ predominantly due to the detector's cartwheel motion. Also, the relative spacecraft velocities $\dot L_d$ lead to minor corrections of the TDI combinations predominantly through relations of the form
\beq
L_{d_1}(t+L_{d_2}(t)/c)\approx L_{d_1}(t)+\dot L_{d_1}(t)\cdot L_{d_2}(t)/c
\label{eqDistSC}
\eeq
with typical relative speeds $\dot L_d\approx 10\cdot e^2 {\rm AU}\cdot f_{\rm orb}\approx 5\cdot10^{-4}\,$m/s (a smaller contribution comes from a term which is proportional to $e^2{\rm AU}^2f^2_{\rm orb}/c$, see appendix A in \cite{Val2005a} for details). The assumption for Eq.~(\ref{eqDistSC}) is that the distance between spacecrafts does not change much during a light travelling time $L/c$. Henceforth, to make our descriptions more readable, we will not make explicit reference to the optical-bench noise. It should be automatically included into the argument whenever we mention laser frequency noise. Algebraically, it is always possible to treat optical-bench noise effectively as additional laser noise.

Previous investigations led to the introduction of three generations of TDI combinations which cancel laser frequency noise based on various assumptions \cite{TEA2004,Val2005a}. The {\it first-generation} combinations are defined to cancel laser noise of a detector which does not have the cartwheel or relative spacecraft motion. If they were used to analyze realistic data, then residual laser noise would contribute to the total instrumental noise. The same is true for the {\it modified first-generation} variables which are based on the assumption that the detector is a rigid object which may perform the cartwheel motion. However, residual laser noise will be much weaker in this case, since it is exclusively caused by the relative motion of spacecrafts which is a second-order effect in terms of the orbital eccentricity $e$. Finally, the {\it second-generation} combinations take relative motions of the spacecrafts into account and cancel noise contributions which depend linearly on $\dot L_d/c\propto e^2$. We claim that choosing second-generation instead of modified first-generation TDI variables has a much weaker influence in the case of BBO than for LISA. The reason is that the relative motion of BBO spacecrafts compared to LISA spacecrafts is a factor $10^4$ smaller. Investigating residual laser noise spectra in modified first-generation and second-generation combinations of BBO will show which generation has to be implemented. However, at least for the purpose of this paper, we just need to introduce the TDI combinations in their modified first-generation form. 

\begin{figure}[ht]
\includegraphics[width=6cm]{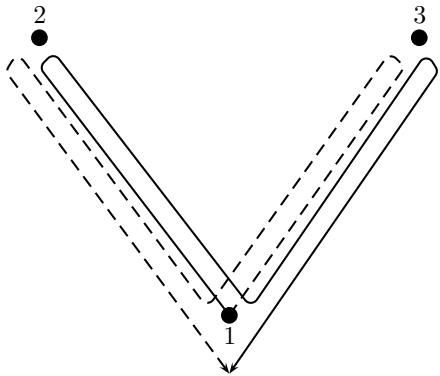}
\caption{Graphical representation of the unequal-arm Michelson TDI combination $X_1$ \cite{Val2005b}. This combination of photo currents mimics the subtraction of two counter-propagating beams.}
\label{figTDIX}
\end{figure}

In order to obtain a concise expression of the TDI combinations, we define new Doppler variables where a certain combination of intra-spacecraft links $z$ is added to inter-spacecraft links $y$
\beq
y_l^\prime\equiv y_l^{\phantom{a}}+\frac{1}{2}(z_{-l,l}^{\phantom{a}}-z_l^{\phantom{a}})
\eeq
In terms of these quantities, the laser-noise-free TDI combination $X_1$ can be cast into the form \cite{VCT2008}
\beq
\begin{split}
X_1(t)&\equiv[y^\prime_{-3,32-2}+y^\prime_{3,2-2}+y^\prime_{2,-2}+y^\prime_{-2}]\\
&\quad-[y^\prime_{2,-2-33}+y^\prime_{-2,-33}+y^\prime_{-3,3}+y^\prime_{3}]
\end{split}
\label{eqTDIX1}
\eeq 
Time delays commute in first-generation variables and therefore, semicolons in Eq.~(\ref{eqDelay}) have been substituted by commas. TDI $X_1$ mimics an unequal-arm Michelson interferometer centered at spacecraft 1. Cyclic permutation of all indices leads to the definition of $X_2$ and $X_3$ which represent interferometers centered at spacecrafts 2 and 3. Each of the two square brackets in Eq.~(\ref{eqTDIX1}) comprise terms which represent a complete round trip of light in clockwise and counter-clockwise direction. These two beams are then subtracted from each other to form the unequal-arm Michelson which can be represented geometrically as shown in Fig.~\ref{figTDIX}.
The instrumental noise of the three channels $X_i$ is correlated. It is more convenient to use channels with uncorrelated instrumental noise, especially if information from all channels is combined to provide optimal sensitivity with respect to GWs. These channels are known by the names $A,\,E,\,T$ and can be defined in terms of the basis vectors $X_i$
\beq
\begin{split}
A&=\frac{X_3-X_1}{\sqrt{2}}\\
E&=\frac{X_1-2X_2+X_3}{\sqrt{6}}\\
T&=\frac{X_1+X_2+X_3}{\sqrt{3}}
\end{split}
\eeq
Each of these variables can be seen as one detector and so in principle, each LISA-type detector has to be treated as a network which consists of three independent detectors. It turns out that these channels have quite different sensitivities to GWs and also, correlation measurements between them does not yield the same profit than what one may naively expect from a detector network. We will come back to this in a later section.

\section{Overview of the simulation}
\label{secSimulate}
\begin{figure}[ht!]
\centerline{\includegraphics[width=8cm]{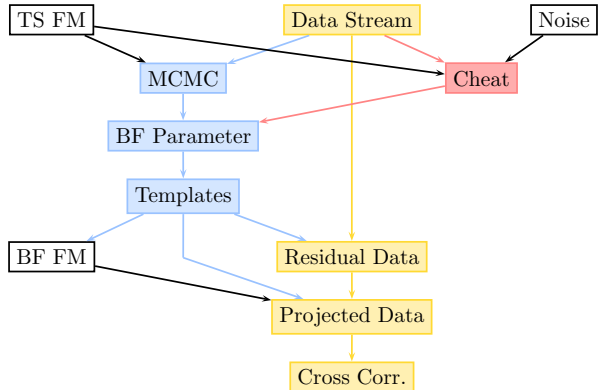}}
\caption{(Color online) FM: Fisher Matrix, TS: True Signal, BF: Best-Fit}
\label{figPipe}
\end{figure}

The simulation is organized according to the pipeline shown in Fig.~\ref{figPipe}. The first step is to generate time series for the various Doppler streams (12 per detector, 48 in total). These data contain the instrumental noise and contributions from 100 CBCs. It is fairly simple to derive time domain models of the test-mass noise ($S_{\rm tm}\propto 1/f^2$ in units of Doppler shift) and the shot noise ($S_{\rm shot}\propto f^2$) \cite{Fra1965}. For the GW signal, we use a time-domain post-Newtonian (pN) approximation (Eq.~(\ref{eqGWform})). The data will depend on spacecraft motion and is generated consistently throughout the network by evaluating the GW phase at retarded time (Eq.~(\ref{eqDoppWave})).
In contrast, the CGWB has to be generated directly in the frequency domain (Eq.~(\ref{eqn:CGWBgen})). Assuming a Gaussian model, the frequency-domain representation of the CGWB is completely determined by a function called the overlap-reduction function which essentially characterizes correlations between different output channels of the detector network (see section \ref{secORF}).

The second step is to search the data for the CBCs. For that purpose, one has to exploit features of the network in a distinct order. The CBCs are analyzed by coherently combining the data of all 12 independent channels. The detector arrangement significantly improves parameter estimation of signals which cannot be integrated over long enough times. Any poorly fitted broad-band signal could have a devastating effect on the mission goal: it may be that the respective residual noise even after applying the noise projection method is stronger than the CGWB spectrum. Initially, we implemented a Markov-chain Monte-Carlo (MCMC) algorithm which searched for the maximum of the posterior distribution determined by our signal models and simulated data \cite{CoPo2007,CrCo2007}. However, our computational resources were not sufficient to perform a realistic search. Therefore, we decided to {\it calculate} the best-fit parameter values. This is a trick (the "Cheat" box in Fig.~\ref{figPipe}) to avoid the CBC analysis. The idea is to calculate the noise vector on the template manifold which points from the true signal to the best fit. Vectors are defined in tangent spaces, so the best fit has to lie in close proximity to the true signal (high signal-to-noise ratio). Further details can be found in sections \ref{secNumericFisher} and \ref{secMaxLike}. The best-fit waveforms are subtracted from the data which then consists of instrumental noise, the CGWB and the subtraction noise.

The third step is to carry out the projection method to remove the subtraction noise. The projection operator is defined in section \ref{secProjection}. This step is required for the following reason. A final correlation measurement of data streams of the two collocated detectors is supposed to lift the CGWB which is correlated to some degree in different channels above any other contribution to the data. Now, this only works if correlations between channels of the instrumental noise and the subtraction noise can be neglected. This is true for the instrumental noise, but it is not for the subtraction noise which is highly correlated. Remember, the subtraction noise corresponds to the difference of the true signal and the estimated signal which is a single quantity for the whole network (modulo detector transfer functions). 

Finally, as described in section \ref{secCross}, we use cross-correlation results to obtain a signal-to-noise ratio (SNR) for the CGWB with a given energy density. Knowing how the SNR scales with observation time, we derive a prediction for BBO's sensitivity towards the CGWB based on its full mission lifetime.

\section{The Fisher matrix in a TDI framework}
\label{secFisher}
\subsection{The signal model and its derivatives}
Since decades, people have been developing a geometrical interpretation of data analysis. These models usually consist of a distribution carried by a certain model manifold \cite{Efr1975,Ama1982,Mit1988}. If one considers Gaussian distributions, the metric of the statistical manifold is given by the Fisher-information matrix $\Gamma^{\phantom{a}}_{\alpha\beta}$. To calculate it, one needs a noise model and a signal model. The model ${\mathcal T}^{\rm m}$ which determines the signal inside a TDI combination $\mathcal T$ depends on parameters $\lambda^\alpha$. Concerning the noise model, one assumes complete knowledge of its (double-sided) spectral density $S^{\rm n}$. In practice, the model for the noise spectral density itself would depend on a few parameters which would have to be estimated before searching the data for certain signals. 

In general, the Fisher matrix $\Gamma^{\phantom{a}}_{\alpha\beta}$ is associated with a TDI variable $\mathcal T$ and it is defined as a scalar product of derivatives of the signal model $\mathcal T^{\rm m}$ with respect to the model parameters $\vec\lambda$. Defining $\partial^{\phantom{a}}_\alpha \mathcal T^{\rm m}\equiv\partial \mathcal T^{\rm m}/\partial\lambda^\alpha$, the Fisher matrix assumes the form
\beq
\begin{split}
\Gamma^{\phantom{a}}_{\alpha\beta}&=\langle\partial^{\phantom{a}}_\alpha \mathcal T^{\rm m}|\partial^{\phantom{a}}_\beta \mathcal T^{\rm m}\rangle\\
&\equiv2\int\limits_0^\infty\drm f\frac{{\rm Re}(\partial^{\phantom{a}}_\alpha \tilde {\mathcal T}^{\rm m}(f)\partial^{\phantom{a}}_\beta \tilde {\mathcal T}^{\rm m*}(f))}{S^{\rm n}(f)}
\end{split}
\label{eqFisher}
\eeq 
The model which determines a single CBC signal inside our simulation depends on 5 extrinsic parameters ($\lambda^0$ = luminosity distance $r$ of the source to the detector, $\lambda^1$ = declination $\theta$, $\lambda^2$ = right ascension $\phi$, $\lambda^3$ = polarization angle $\psi$, $\lambda^4$ = inclination angle $\iota$ of binary orbit with respect to line-of-sight) and 4 intrinsic parameters ($\lambda^5$ = orbital phase $\phi_{\rm c}$, $\lambda^6$ = coalescence time $t_{\rm c}$, $\lambda^7$ = total mass $M=M_1+M_2$ of the binary system, $\lambda^8$ = reduced mass $\mu=M_1M_2/(M_1+M_2)$). In other words, we neglect the spin of the two binaries and assume zero eccentricity of their orbit. If two signals have to be parameterized, then $\lambda^9$ would be the distance parameter of the second signal etc. A simulation is completely based on signal models (there is no true data). Still, we have to distinguish between the model for the simulated signal $\mathcal T^{\rm h}$ which adds to the noise $\mathcal T^{\rm n}$ to form the total data $\mathcal T^{\rm s}$, and the model $\mathcal T^{\rm m}$ which is used to analyze the data. In general, one may choose different models to generate and analyze data. Once generated, $\mathcal T^{\rm h}$ is not considered as a function which depends on parameters, but as a set of fixed numbers. Here, we use the same model to generate and analyze data, and therefore the next paragraph gives a description of both.

Since Eq.~(\ref{eqTDIX1}) tells us that a TDI variable is a linear combination of the Doppler signals $y_l^{\phantom{c}}$, derivatives of a TDI variable with respect to certain model parameters can be expressed as a sum of derivatives of our signal model $y_l^{\rm m}$ for a single Doppler signal. Therefore, it suffices to calculate and present the derivatives $\partial^{\phantom{a}}_\alpha y_l^{\rm m}$. The Doppler signal $y_l^{\rm m}$ is a projection of the GW tensor onto the light-travelling direction $\hat n_l$. In transverse-traceless coordinates, the matrix representation of the GW tensor contains the two GW polarizations $h_+,\,h_\times$ which are functions of the distance $r$ and all intrinsic parameters:
\beq
\begin{split}
h_+(t)&=\frac{c}{2r}\left[5\frac{(\mathcal{M}_{\rm c})^5}{t_{\rm c}-t}\right]^{1/4}\cdot (1+\cos^2(\mathcal{\iota}))\cdot\cos(\phi(t)+\phi_{\rm c})\\
h_\times(t)&=\frac{c}{2r}\left[5\frac{(\mathcal{M}_{\rm c})^5}{t_{\rm c}-t}\right]^{1/4}\cdot 2\cos(\mathcal{\iota})\cdot\sin(\phi(t)+\phi_{\rm c})
\end{split}
\label{eqGWform}
\eeq
We implement the restricted waveform which neglects all harmonics higher than twice the orbital frequency and whose amplitude is determined by the chirp mass $\mathcal M_{\rm c}\equiv GM^{2/5}\mu^{3/5}/c^3$. The evolution of the GW phase is given by a 3.5 post-Newtonian expansion
\beq
\phi(t)=-\frac{2}{\eta}\sum\limits_{k=0}^7p_k\tau^{(5-k)/8}
\label{eqPNphi}
\eeq
with $\tau\equiv(\eta c^3(t_{\rm c} - t))/(5GM)$ and expansion coefficients \cite{BlEA2002}
\vspace*{0.4cm}
\begin{widetext}
\beq
\begin{split}
p_0 & = 1\\
p_1 & = 0\\
p_2 & = \frac{3715}{8064}+\frac{55}{96}\eta\\
p_3 & = -\frac{3}{4}\pi\\
p_4 & = \frac{9275495}{14450688}+\frac{284875}{258048}\eta+\frac{1855}{2048}\eta^2\\
p_5 & = \left(-\frac{38645}{172032}-\frac{15}{2048}\eta\right)\pi\log\left(\frac{\tau(t)}{\tau(0)}\right)\\
p_6 & = \frac{831032450749357}{57682522275840}-\frac{53}{40}\pi^2-\frac{107}{56}C+\frac{107}{448}\log\left(\frac{\tau(t)}{256}\right) +\left(-\frac{126510089885}{4161798144}+\frac{2255}{2048}\pi^2\right)\eta+\frac{154565}{1835008}\eta^2-\frac{1179625}{1769472}\eta^3\\
p_7 & = \left(\frac{188516689}{173408256}+\frac{140495}{114688}\eta-\frac{122659}{516096}\eta^2\right)\pi
\end{split}
\eeq
\end{widetext}
which are most suitably expressed in terms of the symmetric mass ratio $\eta\equiv\mu/M$, and $C=0.57721566\ldots$ is Euler's constant. In total, the GW phase depends on the mass parameters $M,\,\mu$ and the chirp time $t_{\rm c}$. At some point we had to choose a convenient mass parametrization. Obviously, it would have been possible to use $\eta$ instead of $\mu$, and indeed in many situations this could be a good choice. However, for comparable mass binaries like neutron-star binary systems (which is the only kind of signal we included in our simulation), the mass ratio $\eta$ has the odd property to be close to its maximum value $\eta_{\rm max}=0.25$ which holds for equal-mass binaries. By consequence, in our case, probability distributions for $\eta$ will not be Gaussian and the distribution of other parameters may also exhibit non-Gaussian features through parameter correlations. So, without further investigations we decided to use the reduced mass $\mu$ as second mass parameter. As we will show later, the distributions for $M$ and $\mu$ are highly correlated even for strong signals which complicates the calculation of the inverse of the Fisher matrix. It would be interesting to investigate whether a different mass parametrization behaved better in this respect.

Now, projecting the GW tensor, we arrive at the following form for the Doppler signal \cite{CoRu2003}:
\beq
\begin{split}
y_l^{\rm m}(t)=&\frac{1}{2(1-\vec k(\theta,\phi)\cdot\vec n_l(t))}\cdot\vec n^{\top}_l(t)\\
&\cdot\bigg(\sum\limits_{I=+,\times}{\bf e}_I(\theta,\phi,\psi)\cdot h_I(t_{\rm s}-\vec k(\theta,\phi)\cdot \vec r_{\rm s}(t_{\rm s})/c)\\
&\qquad-\sum\limits_{I=+,\times}{\bf e}_I(\theta,\phi,\psi)\cdot h_I(t-\vec k(\theta,\phi)\cdot \vec r_{\rm r}(t)/c)\bigg)\\
&\cdot\vec n_l(t)
\end{split}
\label{eqDoppWave}
\eeq
The Doppler signal $y_l^{\rm m}$ is the GW induced frequency change of light which is sent at $t_{\rm s}=t-L_l(t)/c$ from a spacecraft at position $\vec r_{\rm s}(t_{\rm s})$ and received at time $t$ by its neighbor at position $\vec r_{\rm r}(t)$. ''$\top$'' denotes a transposition. The two polarization matrices ${\bf e}_+,\,{\bf e}_\times$ are derived from their simple form in the GW propagation frame by a rotation $\mathcal{D}(\theta,\,\phi,\,\psi)$ into the solar barycentric frame whose coordinates are used to describe the spacecraft positions. The rotation matrix is shown explicitly in the appendix of \cite{Val2005a}. The propagation direction of the GW is given by $\vec k=-(\cos(\theta)\cos(\phi),\cos(\theta)\sin(\phi),\sin(\theta))$. To calculate the Fisher matrix, we generate time series of derivatives $\partial^{\phantom{a}}_\alpha \mathcal T^{\rm m}$ with $\mathcal T\in\{A,E,T\}$ and subsequently apply a fast-Fourier transform (FFT) to obtain the amplitudes which govern the Fisher matrix components. Taking derivatives with respect to $r$ and $\phi_{\rm c}$ is trivial. The same is true for all other extrinsic parameters, although the result is rather complicated due to the waveform's dependence on the angles $\theta,\,\phi,\,\psi$ (see Eq.~(\ref{eqDoppWave})). However, we think that it is instructive to present derivatives of the GW phase, since for all three non-zero derivatives of the phase, the result can be cast into a form which resembles the pN expansion Eq.~(\ref{eqPNphi}) of the phase. The corresponding expressions can be found in appendix \ref{secAppDer}. 

Finally, one has to specify models for the noise spectral densities $S^{\rm n}(\mathcal T)$. The spectral densities for the uncorrelated channels read \cite{VCT2008}
\beq
\begin{split}
S^{\rm n}(A)&=S^{\rm n}(E)\\
&=16\sin^2(2\pi fL/c)\\
&\qquad\qquad\cdot(3+2\cos(2\pi fL/c)+\cos(4\pi fL/c))S^{\rm tm}\\
&\quad+8\sin^2(2\pi fL/c)(2+\cos(2\pi fL/c))S^{\rm shot}\\[0.2cm]
S^{\rm n}(T)&= 128\sin^2(2\pi fL/c)\sin^4(\pi fL/c)S^{\rm tm}\\
&\quad+16(1-\cos(2\pi fL/c))\sin^2(2\pi fL/c)S^{\rm shot}
\end{split}
\label{eqNoise}
\eeq
with test-mass noise $S^{\rm tm} = S^{\rm acc}/(2\pi fc)^2$ and shot noise $S^{\rm shot} = \hbar\omega_0/P_{\rm rec}(2\pi f/\omega_0)^2$ in terms of double-sided spectral densities. The standard design of BBO provides a spectral density of test-mass acceleration $S^{\rm acc}=9\cdot10^{-34}\,\rm m^2/s^4/Hz$ which in our simulation is assumed to be equal for all test masses and light power $P_{\rm rec}=9\,$W which is received by a spacecraft from one of its neighbors. The carrier frequency of the laser is $\omega_0=5.31\cdot 10^{15}\,\rm s^{-1}$. It is sufficient to express the noise models in terms of the nominal arm length $L=50000\,$km, because estimation errors of the noise will probably exceed the systematic errors due to implementing a simplified model. This is certainly true in our simulation where an estimation of the low-frequency test-mass noise spectrum would be based on a few frequency bins. In contrast, the simulated noise $\mathcal T^{\rm n}$ is based on a combination of individual Doppler signals which then depends on detector motion and asymmetry. 

\subsection{Numerical evaluation of Fisher matrices}
\label{secNumericFisher}
\begin{figure}[ht!]
\centerline{\includegraphics[width=8cm]{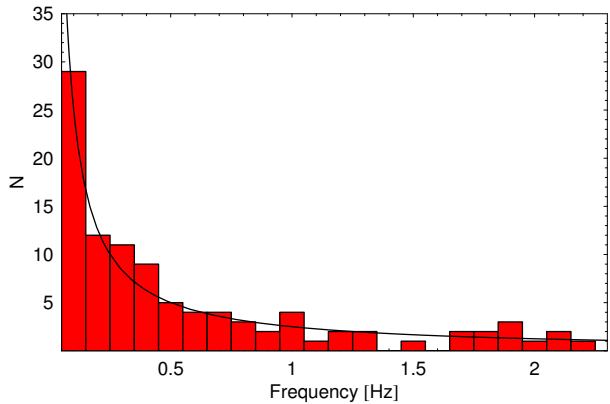}}
\caption{(Color online) Initial distribution of NS/NS signals over frequency bins.}
\label{figNSNSDis}
\end{figure}

\begin{figure}[ht!]
\centerline{\includegraphics[width=6cm,angle=-90]{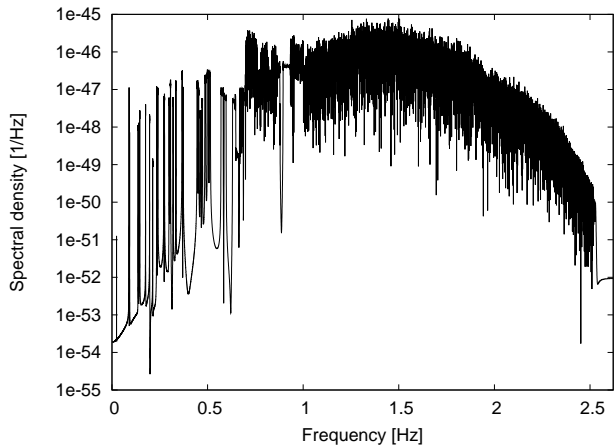}}
\caption{Signal spectrum of the A channel of the first detector. At low frequencies up to 0.7\,Hz, the spectrum features distinguishable, mildly chirping signals. At high frequencies, many signals overlap to form a plateau of chirps. The signals' merger times were chosen such that, within $T=10^5\,$s, signal frequencies never become greater than 2.6\,Hz, which is half the sampling frequency. Note that at frequencies above 0.5\,Hz, just every 20th frequency bin is plotted to reduce figure size.}
\label{figSigSpec}
\end{figure}

The complete simulated CBC foreground is composed of 100 NS/NS systems which occupy frequencies between $52\,$mHz and $2.2$Hz. Restricted by computational power of single notebooks (a cluster version of the code is being developed), we could simulate data with an observation time $T=10^5\,$s and a sampling frequency of $f_{\rm s}=5.24288\,$Hz which essentially fix the frequency range of the injected binaries. Parameter values for $t_{\rm c}$, $r$ and the two mass parameters of the CBCs are drawn from non-uniform priors. Values for $M$ and $\mu$ are derived from normal distributions of the individual masses $M_1,\,M_2$ which are centered at $1.4M_\odot$, distance values are restricted to yield sensible signal-to-noise ratios and values for the chirp time are determined by assuming a certain distribution of signals over frequency bins. Assuming a Newtonian evolution of the orbital frequency of the binaries,  the number of signals per frequency bin has to obey a distribution $N(f)\propto 1/f^{11/3}$ near BBO frequencies \cite{UnVe2001}. However, in our simulation, we draw initial frequencies from a $N(f)\propto 1/f$ distribution which yields a few systems at higher frequencies, but otherwise has no significant effect on our analysis. The frequency distribution of the 100 NS/NS signals is displayed in Fig.~\ref{figNSNSDis}.
The reason for taking greater care in frequency priors is that the signal distribution in frequency space has a significant impact on correlation values between parameter distributions of different CBCs especially since the sky resolution of the detector network is comparatively poor for short observation times. The systems with highest frequencies have chirp times $t_{\rm c}$ which are of the order of a few $T$ and therefore the signal spectrum which is shown in Fig.~\ref{figSigSpec} exhibits multiple quasi-monochromatic peaks at low frequencies and a few chirp ''plateaux'' at higher frequencies. 

Before being able to numerically evaluate the Fisher matrix, one has to search the simulated data for the injected binaries and estimate their parameter values. The signal derivatives in Eq.~(\ref{eqFisher}) have to be evaluated at the estimated parameter values $\hat{\vec\lambda}$. Implementing uniform priors of parameter distributions at this point, an optimal analysis is performed by searching the likelihood function $\mathcal L(\vec\lambda)\propto\exp(-1/2\sum_i\langle{\mathcal T}^{\rm s}_i- {\mathcal T}_i^{\rm m}(\vec\lambda)|{\mathcal T}_i^{\rm s}-{\mathcal T}_i^{\rm m}(\vec\lambda)\rangle)$ for its global maximum \cite{Fin1992,CuFl1994}. Here the sum has to be taken over all independent network channels (the BBO network furnishes 12 independent channels, 3 per detector). As was argued in \cite{CuHa2006}, no existing computer or network of computers could accomplish that search for a realistic foreground formed by $10^5-10^6$ CBCs. Even searching simulated BBO data for 100 signals including the estimation of parameters is a difficult task which optimally requires a high-end cluster. Our work is not intended to make any propositions how to perform that search let alone to carry it out. So we have to work around the problem. The idea is that knowing the realization of the instrumental noise in a simulation run -- which we do, since we generate the noise $\mathcal T^{\rm n}$ and add it to the signals $\mathcal T^{\rm h}$ -- and assuming Gaussian distributions for the signal parameters with small estimation errors (high SNR), one can use the following equation to calculate the estimation errors $\delta\lambda^\alpha$ (the difference between the maximum likelihood values $\hat\lambda^\alpha$ and the true parameter values $\lambda^\alpha$ of the signals) \cite{CuHa2006,Val2007}:
\beq
\delta\lambda^\alpha=\Gamma^{\alpha\mu}\sum_i\langle\mathcal T_i^{\rm n}|\partial_\mu \mathcal T_i^{\rm m}\rangle
\label{eqParError}
\eeq
Again, the sum has to be taken over all independent channels $\mathcal T_i$ of the detector network. The parameter errors depend on the inverse of the network Fisher matrix $(\Gamma^{\alpha\beta})\equiv(\sum_i\Gamma^i_{\alpha\beta})^{-1}$ with $\Gamma^i_{\alpha\beta}\equiv\langle\partial_\alpha\mathcal T^{\rm m}_i|\partial_\beta\mathcal T^{\rm m}_i\rangle$. The parameter estimation errors are added to the true parameter values of the signals injected into the simulation pipeline, and the Fisher matrix can finally be evaluated. A brief introduction into the geometric interpretation of the Fisher matrix and how to make use of it can be found in section \ref{secProj} which also explains Eq.~(\ref{eqParError}). The reader may have worked with a close relative of Eq.~(\ref{eqParError}) in another context. It is a generalization of the ${\mathcal F}$-statistic equation to obtain best fits of its 4 amplitude parameters \cite{CuSc2005}. The ${\mathcal F}$-statistic is based on templates which are linearized with respect to $r,\,\iota,\,\psi$ and $\phi_c$. It is straightforward to show that if one substitutes the complete data $\mathcal T_i^{\rm s}$ for the noise $\mathcal T_i^{\rm n}$ in Eq.~(\ref{eqParError}), then the equation directly yields the best fit of any parameter which enters linearly into the definition of the template. Our model does not have linear parameters, but many alternative template models do have.

Notice that by calculating the best fits, we neglect the detection problem, i.e.~we assume that all binaries in the data are detected. We should also mention that our method to calculate the best fits represents an optimal analysis scheme. The optimal scheme is to search simultaneously for all signals. A more realistic search which requires much less computational power is the hierarchical search. There, one detects signals one by one, starting with highest SNR and ''digging'' down the signal spectrum until the last binary is identified and measured. Unresolved binaries act as confusion noise. Identified binaries are subtracted from the data so that during the hierarchical search, parameters of already detected binaries are constantly refined as the confusion noise decreases. This scheme has been studied by means of a self-consistent recursive evaluation in \cite{CuHa2006}. In contrast, the optimal search is not corrupted by confusion noise. Correlations between different signals, which lead to confusion noise in the hierarchical search, are incorporated into the signal model of the optimal search. The only possible shortcoming of an optimal search is that it may fail to accurately estimate parameter values of the model (including to find the right template-manifold dimension which depends on the number of detected signals). 

\subsection{Multi-signal templates and Fisher-matrix inversion}
\label{secMultiSig}
Given 100 signals which each depend on 9 parameters, the total Fisher matrix $\Gamma^i_{\alpha\beta}$ for each channel $i$ becomes a 900$\times$900 matrix. It turns out that inverting the Fisher matrix is a highly nontrivial task. In our simulation, inverse Fisher matrices are used in two different ways. First, we need the network matrix evaluated at the true parameter values to compute the estimation errors by means of Eq.~(\ref{eqParError}). Second, the inverse Fisher matrices of individual channels evaluated at the estimated parameter values have to be computed to define the subtraction-noise projector in Eq.~(\ref{eqProjOP}). To start with, we outline a generic inversion scheme which, in the end, does not solve all problems. However, this method still forms the foundation of the complete solution. In the next part of this section, we omit the channel index $i$, since the described method is used in the same way to invert channel and network Fisher matrices.

The inversion procedure starts with the computation of a new matrix $\Gamma^\prime_{\alpha\beta}=\Gamma^{\phantom{a}}_{\alpha\beta}/\sqrt{\Gamma^{\phantom{a}}_{\alpha\alpha}\Gamma^{\phantom{a}}_{\beta\beta}}$ such that $\Gamma^\prime_{\alpha\alpha}=1$ and all off-diagonal components have an absolute value smaller than one. This step is necessary since in our simulation the numerical range of Fisher-matrix components is $10^{-50}$ --- $10^5$. By consequence, ratios of different eigenvalues of the Fisher matrix can assume large values. Such a matrix is called {\it ill-conditioned} and is known to be hard to invert numerically, because tiny inaccuracies of a few matrix components may have a great effect on the eigenvalues or the components of the inverted matrix. These inaccuracies are unavoidable due to limited machine precision. It turns out that $\Gamma^{\prime}_{\alpha\beta}$ is still ill-conditioned and cannot be inverted using standard double-precision variables. At this point, one has to implement a multi-precision package into the code. We found that CLN (Class Library for Numbers) \cite{HaKr2008} provides all required functions. Using a 50-digit precision, the scaled matrix can be inverted following its LU decomposition. Next, the inverted matrix is scaled back to form the inverted Fisher matrix $\Gamma^{\alpha\beta}=\Gamma^{\prime\,\alpha\beta}/\sqrt{\Gamma^{\phantom{a}}_{\alpha\alpha}\Gamma^{\phantom{a}}_{\beta\beta}}$. We expect that the degree of ill-conditioning decreases significantly once much longer observation times can be simulated. To explain this, we need to have a look on the correlation matrix of a single binary. The correlation matrix is derived from the covariance matrix $\Gamma^{\alpha\beta}$ in the same way than $\Gamma^\prime_{\alpha\beta}$ was derived from $\Gamma^{\phantom{c}}_{\alpha\beta}$.
\begin{table}[ht!]
\begin{tabular}{c||c|c|c|c|c|c|c|c|c}
$\vec\lambda$ & $r$ & $\theta$ & $\phi$ & $\psi$ & $\iota$ & $\phi_{\rm c}$ & $t_{\rm c}$ & $M$ & $\mu$ \\\hline\hline
$r$ & 1 & -0.287 & -0.038 & -0.049 & -0.999 & -0.066 & -0.066 & 0.050 & -0.048\\\hline
$\theta$ & & 1 & 0.162 & 0.118 & 0.287 & -0.189 & -0.189 & -0.062 & 0.068 \\\hline
$\phi$ & & & 1 & -0.055 & 0.038 & 0.043 & 0.043 & 0.045 & -0.046 \\\hline
$\psi$ & & & & 1 & 0.049 & 0.014 & 0.014 & -0.084 & 0.084 \\\hline
$\iota$ & & & & & 1 & 0.066 & 0.066 & -0.050 & 0.048\\\hline
$\phi_{\rm c}$ & & & & & & 1 & 0.999 & -0.017 & -0.011\\\hline
$t_{\rm c}$ & & & & & & & 1 & -0.014 & -0.014\\\hline
$M$ & & & & & & & & 1 & -0.999\\\hline
$\mu$ & & & & & & & & & 1 
\end{tabular}
\caption{Network correlation matrix for a single NS/NS at 0.57\,Hz. Some correlation coefficients strongly depend on the observation time which is $T=10^5\,$s in this case.}
\label{tabCorr1}
\end{table}
As can be seen in Tab.~\ref{tabCorr1}, correlation is especially strong between $r\leftrightarrow\iota$, $\phi_{\rm c}\leftrightarrow t_{\rm c}$ and $M\leftrightarrow\mu$. Strong correlations indicate that by changing one parameter from its best-fit value, the loss in accuracy of the waveform fit can be compensated by changing the value of the other parameter of the correlation pair. Therefore, at first sight, it seems to be obvious that these pairs may be strongly correlated. The question is, under which circumstances these correlations become weaker. The pair $M\leftrightarrow\mu$ decorrelates once a considerable amount of the chirp is observed and the signal-frequency change accelerates. It is well-known that the low-frequency evolution of CBC waveforms is completely determined by a single mass parameter, the binary's chirp mass $\mathcal M_{\rm c}$. For those waveforms, implementing a model which needs two mass parameters must exhibit maximal correlations between them. A similar argument can be invoked for the pair $\phi_{\rm c}\leftrightarrow t_{\rm c}$. By consequence, correlation matrices of signals with higher frequencies have lower correlation coefficients for these pairs. A decorrelation of $r\leftrightarrow\iota$ (and weakening of many other correlation coefficients) is observed as soon as the orbital motion of the detectors leads to measurable amplitude and phase modulations of the signal. With maximal observation times of $T=10^5\,$s, we are not able to study the impact of the Doppler shift on parameter estimations. However, as we are not particularly interested in the quality of best fits, but accept any quality as long as a projection of subtraction noise can be carried out successfully, we do not investigate parameter correlations further in this paper. It turns out that the best fits are accurate enough for this purpose.

As mentioned in the beginning, the inversion algorithm as presented in the last paragraph does not provide a complete solution of the inversion problem. The reason is that a 900$\times$900 matrix determined by multi-precision components needs too much memory and even if it can be kept in memory during runtime (e.g.~by implementing specifically designed inversion schemes \cite{ImLa1989}), then the inversion would take too much time. To solve this problem, we have to understand a little more about Fisher matrices. Consider a matrix which includes $N$ copies of single CBC templates. Each CBC is determined by $P$ parameters. In our case, the number of templates is $N=100$ and the number of parameters is $P=9$. Let us introduce the ''confused'' Fisher matrix 
\beq
\Gamma^0 = 
\begin{pmatrix}
g^1 & 0 & \ldots & 0 \\
0 & \ddots &  & \vdots \\
\vdots & & & 0 \\
0 & \ldots & 0 & g^N
\end{pmatrix}
\eeq
It is a block matrix which contains the Fisher matrices $g^k_{\alpha\beta}$ with $\alpha,\beta\in\{1,\ldots,P\}$ of $N$ CBCs on its diagonal. In other words, it differs from the total Fisher matrix by neglecting correlations between different signals. We call it confused, because whenever this matrix is applied instead of the total matrix, it is like our knowledge of correlations between different signals is ignored. Correlation coefficients become random variables in the analysis pipeline leading to confusion noise. We claim that it is legitimate to use the block matrix when calculating estimation errors by means of Eq.~(\ref{eqParError}). To support this claim, one has to investigate the impact of the correlation coefficients on the eigenvalues of Fisher matrices of individual CBCs. Fisher matrices (their inverses to be precise) define a multi-variate Gaussian distribution in parameter space. Their eigenvalues correspond to the variances of the distribution along its principal axes. The question is what happens to the distribution defined by a matrix $g^k_{\alpha\beta}$ when the correlations between signal $k$ and other signals are incorporated into the model. This problem can be treated with perturbation theory similar to perturbations of a Hamiltonian (here, $\Gamma^0$) which is weakly perturbed by interactions
\beq
C=
\begin{pmatrix}
0 & g^{12} & \ldots & g^{1N} \\
g^{12,\,\top} & \ddots &  & \vdots \\
\vdots & & & g^{N-1,N} \\
g^{1N,\,\top} & \ldots & g^{N-1,N,\,\top} & 0
\end{pmatrix}
\eeq
where $g_{\alpha\beta}^{ij}$ are the correlation coefficients between signals $i,j\in\{1,\ldots,N\}$ and ''$\top$'' denotes the transposition of a matrix. The proper condition to justify the perturbation approach is that correlations $g^{ij}$ have to be small compared to differences of eigenvalues of $g^i$ and $g^j$, which is the case for any combination of signals in our simulation. For this particular form of perturbation $C$, theory tells us that the $l$th eigenvalue $\xi^0_{kl}$ of the Fisher matrix $g^k$ ($l\in\{1,\ldots,P\}$) is perturbed at second order in $C$ according to
\beq
\xi_{kl}^{\phantom{a}}=\xi_{kl}^0+\sum\limits_{i\neq k}^N\sum\limits_{j=1}^P\frac{\left|\big\langle kl^0\big|g^{ki}_{lj}\big|ij^0\big\rangle\right|^2}{\xi_{kl}^0-\xi_{ij}^0}+\mathcal{O}(C^4)
\eeq
where $|kl^0\rangle$ are the $P$ eigenvectors of the Fisher matrix $g^k$ with eigenvalues $\xi_{kl}^0$. The perturbation of the eigenvectors reads
\beq
|kl\rangle=|kl^0\rangle+\sum\limits_{i\neq k}^N\sum\limits_{j=1}^P\frac{\big\langle kl^0\big|g^{ki}_{lj}\big|ij^0\big\rangle}{\xi_{kl}^0-\xi_{ij}^0}\cdot|ij^0\rangle+\mathcal{O}(C^2)
\label{eqEigenVec}
\eeq
Therefore, up to first order in $C$, one can say that the multi-variate Gaussian does not change the lengths of its major axes. Instead, the distribution is rotated and the small rotation angles are given by the fraction in Eq.~(\ref{eqEigenVec}). This means that, when using the block matrix $\Gamma^0$ instead of the total Fisher matrix $\Gamma$, the coordinate basis $|\partial_\mu \mathcal T^{\rm m}_i\rangle$ in Eq.~(\ref{eqParError}) is misaligned with respect to the inverted Fisher matrix $(\Gamma^0)^{-1}$, and that the parameter errors $\delta\lambda^\alpha$ lie in false ''directions'', but giving rise to a comparable accuracy of the waveform fit up to order $\mathcal O(C)$. That is the reason why we may use the block matrix at this point. The projection method described later does not depend on these rotations of the parameter space. The benefit is that we can easily invert the confused Fisher matrix by inverting Fisher matrices of each signal. In principle, we could even correct the misalignment by rotating the inverted matrix with rotation angles $\big\langle kl^0\big|g^{ki}_{lj}\big|ij^0\big\rangle/(\xi_{kl}^0-\xi_{ij}^0)$ provided that we also calculate the eigenvalues and eigenvectors of all Fisher matrices $g^k$. As a corollary, we add that correlations between different signals always lead to a loss of Fisher information represented by the determinant of the Fisher matrix
\beq
\det\Gamma=\det\Gamma^0\left(1-\sum\limits_k^N\sum\limits_{i\neq k}^N\sum\limits_{l,j=1}^P\frac{\left|\big\langle kl^0\big|g^{ki}_{lj}\big|ij^0\big\rangle\right|^2}{\xi_{kl}^0\xi_{ij}^0}\right)+\mathcal O(C^4)
\eeq
The second term in round brackets is always positive and well defined, because Fisher matrices are positive definite. So, the decrease of the determinant is a second-order effect in correlation coefficients which is further suppressed by the Fisher information of particular template parameters (i.e.~the respective eigenvalue $\xi_{kl}^0$) and therefore, especially in the high SNR regime, one may neglect information loss.

Unfortunately, we cannot make use of the same simplification when dealing with Fisher matrices which define the projection operator. There, directions reflect the amount of correlations between the template derivatives and actual subtraction noise in the data. These very correlations have to be exploited to facilitate removal of the subtraction noise. Our strategy in this case is to reduce the dimension of the template manifold by projecting a subset of all signals. Namely, we project all signals which possess power in the frequency range which contributes most of the SNR to the final correlation measurements of the CGWB. More details can be found in section \ref{secResProj}.

\section{The stochastic background}
\label{secStochastic}
In this section we sketch out how to simulate the CGWB. The stochastic data is generated directly in the frequency domain starting with one channel and then taking correlations between channels into account to derive data for other channels. The function which describes the correlations is called the overlap-reduction function (ORF) $\gamma_{ab}(f)$ between channels $a$ and $b$ of the same detector or different detectors. As we are going to learn in section \ref{secORF}, correlations between channels A, E, T of the same detector are negligible. We introduce a new definition for the ORF which does not make any attempt to factor out a channel's transfer function. This is the most convenient approach based on a dynamical detector model where relative detector and satellite motion has to be taken into account. Henceforth, since the T channel does not furnish significant sensitivity with respect to a measurement of the CGWB (see \cite{PrEA2002} and section \ref{secORF}), it will be excluded from the branch of the pipeline which processes the CGWB. 

\subsection{Generation of stochastic backgrounds in detector networks}
A zero-mean Gaussian background is completely characterized by its second-order moments, i.e.~the auto- and cross-correlations of TDI channels. In our case, we just have to include channels of the two collocated detectors in our investigations since correlations between other detectors are negligible (correlations fall off rapidly if the distance between detectors becomes much larger than the length of the GW). Therefore, based on correlation properties of a simulated stochastic background signal ${\mathcal T}^{\rm b}$ in detector channels $a$ and $b$, notably
\beq
\frac{1}{T}{\rm Re}\left\{\left\langle\tilde{\mathcal T}_a^{\rm b}(f)[\tilde{\mathcal T}_b^{\rm b}(f)]^*\right\rangle\right\} = S^{\rm b}(f)\gamma_{ab}(f),
\label{eqCGWBCorr}
\eeq
we have to find an algorithm to calculate the stochastic TDI signals ${\mathcal T}^{\rm b}(f)$ produced by all channels of the two collocated detectors. We define the ORF $\gamma_{ab}$ as a real-valued function, which essentially establishes a convention how correlations between channels are evaluated (see Eq.~(\ref{eqCorrelate})). The ORF governs the strength of cross correlations ($a\neq b$) and autocorrelations ($a=b$) in the frequency domain. Here, the observation time $T$ is used to convert amplitude squares into spectral densities and the (double-sided) background spectral density $S^{\rm b}(f)$ of the GW amplitude is related to the fractional energy density $\Omega$ by
\begin{equation}
 S^{\rm b}(f) = \frac{3H_0^2}{4\pi^2} \cdot \frac{\Omega(f)}{f^3}
\end{equation}
The background is assumed to be isotropic and to have a white energy spectrum with fiducial value $\Omega(f) = 10^{-15}$. In our simulation, the value of the Hubble constant is $H_0=72\,\rm km/s/Mpc$. 

Now, the idea is to generate a stochastic signal with the correct spectrum in one channel and then to proceed with other channels by taking mutual correlations into account. The equations used to generate the background are \cite{LSC2007a}
\begin{widetext}
\beq
\begin{split}
 &\T{\mathcal T}_1(f) =  \frac{1}{\sqrt{2}}\sqrt{S^{\rm b}(f) T} \sqrt{\gamma_{11}(f)}(a_1(f) + \irm b_1(f)),\\
 &\T{\mathcal T}_2(f) = \T{\mathcal T}_1(f) \frac{\gamma_{12}(f)}{\gamma_{11}(f)} + \frac{1}{\sqrt{2}} \sqrt{S^{\rm b}(f)T \left(\gamma_{22}(f) - \frac{\gamma_{12}^2(f)}{\gamma_{11}(f)}\right)}(a_2(f) + \irm b_2(f))
 \label{eqn:CGWBgen}
\end{split}
\eeq
\end{widetext}
For each frequency, random values $a_1(f), a_2(f), b_1(f)$ and $b_2(f)$ are drawn from a normal distribution $\mathcal{N}(0,1)$. It is not necessary to extend Eq.~(\ref{eqn:CGWBgen}) to three or more channels, because in the end correlations are evaluated between independent pairs $A_0\leftrightarrow A_1$ and $E_0\leftrightarrow E_1$ of the two collocated detectors. In the next section, we show how to obtain the overlap-reduction function $\gamma_{ab}(f)$ in a TDI network. 

\subsection{Overlap-reduction function}
\label{secORF}
As already mentioned, we need a procedure to calculate the overlap-reduction function (ORF) between arbitrary detector channels at all frequencies. In previous publications, the ORF has been defined by \cite{Fl1993}
\begin{equation}
 \gamma^{\rm pre}_{ab}(f) = \frac{5}{8\pi} \int_{S^2} d\hat{\Omega} \; \e^{2\pi \irm f \hat{\Omega} \Delta \vec{x}/c} (F_a^+F_b^+ + F_a^{\times}F_b^{\times}).
\label{eqORFpre}
\end{equation}
Here $\Delta \vec{x}$ denotes the separation vector between the detectors, $\hat{\Omega}$ is a unit-length vector on the two-sphere $S^2$ and $F_a^{+,\times}$ are the response functions of detector $a$ to the + or $\times$ polarization. This function is normalized such that its value is unity for two coincident, aligned Michelson detectors with perpendicular arms. The motivation for this definition was to separate the optical properties of the detectors from their geometric properties which determine the response functions. The ORF $\gamma^{\rm pre}_{ab}$ is used to calculate correlations between projected GW amplitudes in two detectors, and then optical transfer functions can be used to derive the correlations of the detector outputs. 

There are a few reasons why the ORF in Eq.~(\ref{eqORFpre}) cannot be used under general circumstances. First, it assumes that the response functions are constant in time. Obviously, this is not the case for space-borne detectors, where test masses which are on individual orbits around the Sun move relative to each other. Also, the separation vector between different detectors does not have to be constant. Beyond the long-wavelength limit which demands that the length of a GW is much larger than the dimension of detectors, it is in any case difficult to agree upon a detector position. In other words, light-travelling times between test masses are neglected in $\gamma^{\rm pre}_{ab}$.  Therefore, we propose a slightly different definition of the ORF which is not normalized and which is a direct representation of the correlation strength between channels. We will not make any attempts to separate optics and geometry, since they are tightly linked in TDI detectors. The complete detector dynamics can be incorporated into a frequency-domain correlation function in the following way:
\begin{enumerate}
\item Inject a time-domain delta signal $\delta(t,t^\prime)=f_{\rm s}\cdot{\rm sinc}(\pi f_{\rm s}(t-t^\prime))$ ($f_{\rm s}$ being the sampling frequency) associated with a polarization and sky direction into each detector of the network. Propagate it from a common origin so that relative phase shifts of the GW at different detectors are automatically taken into account. Make sure that $t^\prime$ is larger than any light travelling times between detectors, otherwise the peak does not appear in all data streams.
\item Record the outputs ${\mathcal T}_a^{\rm w\,+,\times}(\theta,\phi;t)$ of all TDI channels.
\item Apply an FFT to the data and thereby obtain the complex-valued transfer functions $\T{\mathcal T}_a^{\rm w\,+,\times}(\theta,\phi;f)$ of the TDI channels. In our case, the transfer function can be used to map GW amplitudes to TDI Doppler outputs in the frequency domain. 
\item Multiply the transfer functions of different channels and average the product over many sky directions and polarizations to obtain the ORF.
\end{enumerate}
In summary, the ORF for a TDI network is defined by
\beq
 \gamma_{ab}(f) =  \left\langle\sum\limits_{I=+,\times}{\rm Re}\left\{\T{\mathcal T}_a^{{\rm w}\,I}(f,\theta,\varphi)[\T{\mathcal T}_b^{{\rm w}\,I}(f,\theta,\varphi)]^*\right\}\right\rangle_{\rm s.a.}
\label{eqORFnew}
\eeq
In our case the phase factor $\e^{2\pi \irm f \hat{\Omega} \Delta \vec{x}/c}$ arising from the time delay between the detectors is already included in the Doppler signal outputs of the TDI. We found that averaging over 200 random sky directions provides very accurate results. Values for the right ascension $\phi$ are drawn from a uniform distribution $\mathcal{U}(0,2\pi)$ whereas values for the declination $\theta$ are obtained by calculating the $\arcsin$ of values drawn from a uniform distribution $\mathcal{U}(-1,1)$, which entails an isotropic distribution of corresponding sky directions.

\begin{figure}[ht]
\includegraphics[width=6cm,angle=-90]{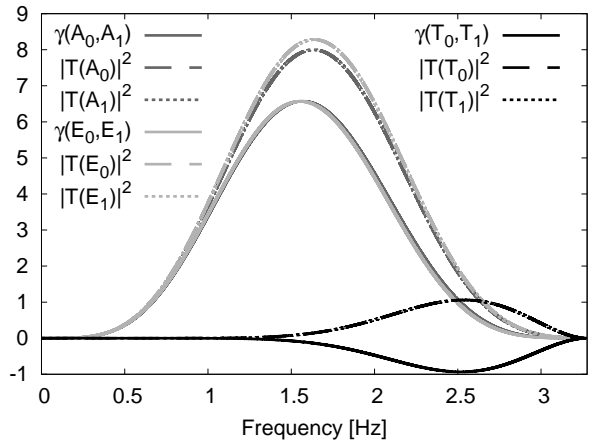}
\caption{The figure shows the overlap-reduction functions $\gamma$ between channels A,E,T of the collocated detectors compared with the corresponding squared transfer functions. The sky average was taken over 400 random sky directions. The squared transfer functions of each channel type are identical in both detectors. At frequencies below 0.1\,Hz all displayed curves related to channels A,E are proportional to $f^4$, the T channel curves are proportional to $f^{10}$! The response of the T channel to GWs is very poor below 2\,Hz.}
\label{figORF-AA}
\end{figure}

The set of curves displayed in Fig.~\ref{figORF-AA} shows the ORFs between channels A,E and T of the two collocated detectors in comparison to the channels'
sky-averaged squared transfer functions (STF).
\begin{equation}
\gamma_{aa}=|T_a(f)|^2 = \left\langle\sum\limits_{I=+,\times}|\T{\mathcal T}_a^{{\rm w}\,I}(f,\theta,\varphi)|^2\right\rangle_{\rm s.a.}
\end{equation}
The response of the T channel lies well below the response of channels A,E at frequencies up to 2\,Hz. More specifically, within the correlation band $0.1$ -- $0.4\,$Hz (see section \ref{secCross}) of BBO, the T channel response (i.e.~expressed as STFs or ORF) is smaller by a factor $\sim2\cdot10^5(0.4\,{\rm Hz}/f)^6$ than the response of the other two channels. This again is the reason why our correlation analysis will not include the T channel.
\begin{figure}[ht]
\includegraphics[width=6cm,angle=-90]{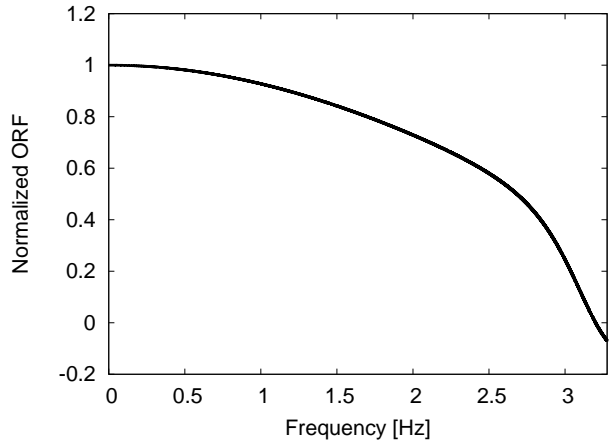}
\caption{The figure displays the normalized ORF between channels $A_0\leftrightarrow A_1$ of the two collocated detectors. The normalized ORF has a maximum at $f=0\,$Hz and then oscillates around zero with constantly decreasing amplitude towards higher frequencies. The ORF as defined in Eq.~(\ref{eqORFnew}) governs the correlation of TDI outputs. In contrast, the normalized ORF is a better representation of the correlation of a stochastic GW signal as input to the detectors. However, one has to keep in mind that all frequency-domain functions are obtained via FFT from a dynamical model and therefore it is not possible in a simple way to deduce GW correlations from measured TDI outputs.}
\label{figORF-norm}
\end{figure}

In this paper, the ORFs are defined as correlation functions between TDI Doppler channels which are -- at low frequencies -- proportional to the square of the second (e.g.~A,E) or even higher derivative (e.g.~T) of the GW amplitude. To find a better measure of stochastic GW background correlations (as projection/combination onto two different TDI channels), one has to compare the ORF with the channel responses to GWs. In Fig.~\ref{figORF-norm}, the ORF between the two A channels of the collocated detectors is shown normalized by the geometrical mean of the two respective STFs:
\beq
\gamma_{A_0A_1}^{\rm N}\equiv \frac{\gamma_{A_0A_1}^{\phantom{N}}}{\sqrt{\gamma_{A_0A_0}^{\phantom{N}}\gamma_{A_1A_1}^{\phantom{N}}}}
\eeq
At frequencies up to 0.5\,Hz, the ORF and the STFs show identical response. This region is called the long-wavelength regime where the length of GWs is much larger than the dimension of the detectors. Beyond the long-wavelength limit, the ORF becomes weaker at higher frequencies compared to the STFs. As we expound in section \ref{secCross}, correlation measurements with BBO detectors will be dominated by contributions of frequencies in the long-wavelength regime.

\begin{figure}[ht]
\includegraphics[width=6cm,angle=-90]{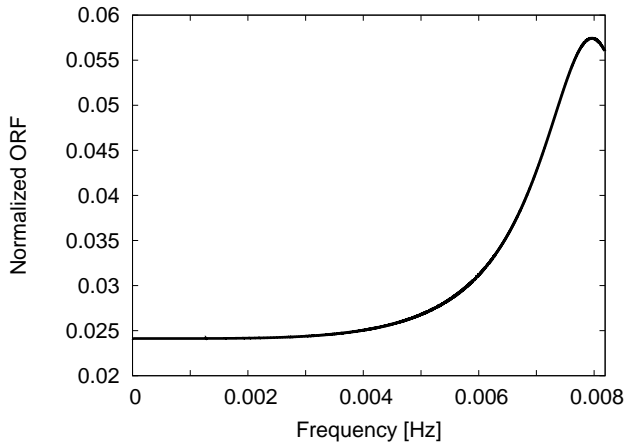}
\caption{The figure shows the normalized ORF between channels $A_0\leftrightarrow E_0$ of the same detector. Correlations between $A_0\leftrightarrow E_0$ are much weaker than correlations $A_0\leftrightarrow A_1$. Still the residual correlation could lead to an increased sensitivity of a single detector to isotropic stochastic backgrounds. Here, the data is based on a model detector with orbital eccentricity $e=0.04$ to make the effect of a residual response due to asymmetries stronger than artifical anisotropies resulting from a sky-average over a finite number of sky directions.}
\label{figORF-AE}
\end{figure}

The graph in Fig.~\ref{figORF-AE} allows to draw another interesting and important conclusion. It shows the normalized ORF between channels A, E of the same detector. Obviously, cross correlating A channels of two different (collocated) detectors provides much more sensitivity than cross correlating independent channels of the same detector. In fact, one can show that if the detectors were equilateral triangles, then the ORF between channels A and E of the same detector would vanish \cite{SeCo2004,Set2006}. Sky-averages of these two channels are orthogonal to each other in terms of their response to an isotropic stochastic GW background, but they do permit non-vanishing correlations arising from higher moments of anisotropic backgrounds (e.g.~the hexadecapole moment). In our simulation, asymmetries in the triangular detector shape are responsible for a residual sky-averaged correlation strength. The detector asymmetry is of the order of the orbital eccentricity of the satellites. To make this effect stronger than artificial anisotropies resulting from a sky average over a finite number of sky directions, we chose a model detector with increased eccentricity value ($e=0.04$) to generate Fig.~\ref{figORF-AE}.
Since BBO has collocated detectors, the residual correlation will not be exploited, because in that case one can form more efficient correlation schemes based on channels of different detectors. Also, when generating the CGWB data, we may neglect correlations between channels A and E which justifies the two-channel approach in Eq.~(\ref{eqn:CGWBgen}) to the BBO network. 

\begin{figure}[ht]
\includegraphics[width=6cm,angle=-90]{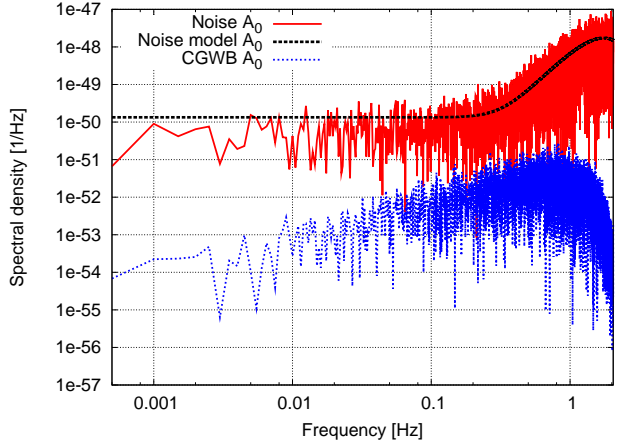}
\caption{Simulated CGWB with fractional energy density $\Omega=10^{-15}$ in comparison to the instrumental noise. A smaller observation time, $T=2\cdot10^3\,$s, was chosen to generate the curve, but the spectra shown are independent of the observation time. The noise model is exclusively used to compute Fisher matrices, projection operators and the SNR (see next section).}
\label{figPrimordial}
\end{figure}

In terms of the calculated ORFs $\gamma_{A_0A_1}(f)$, $\gamma_{E_0E_1}(f)$ and STFs $\gamma_{A_0A_0}(f)$, $\gamma_{A_1A_1}(f)$, $\gamma_{E_0E_0}(f)$ and $\gamma_{E_1E_1}(f)$, we are able to calculate the stochastic background $A^{\rm b}_{0,1}(f)$ and $E^{\rm b}_{0,1}(f)$ by means of Eq.~(\ref{eqn:CGWBgen}). In Fig.~\ref{figPrimordial}, its spectral density in channel $A_0$ is shown together with the instrumental-noise spectral density. One can directly infer from the graph that the $\rm SNR$ is about $0.1$. A correlation measurement has to raise this value to 5 at least. By simple arguments, we can determine the conditions under which the CGWB becomes detectable by means of correlation measurements.
As a first approximation one can say that the correlation measurement effectively shifts the CGWB curve in Fig.~\ref{figPrimordial} upwards by a factor $\sqrt{T}$. Adding the SNRs obtained from two independent correlation pairs, we find that an CGWB with energy density $\Omega=10^{-15}$ can be detected if correlation times exceed $10^4\,$s. Section \ref{secCross} provides more details of the correlation measurement, and we are going to show in section \ref{secResults} that our first guess gives the right order of magnitude for the minimal correlation time.

\section{Template-based projections of subtraction noise}
\label{secProj}

In this section we introduce a differential geometric point of view based in principal on the works of S.-I.~Amari \cite{Ama1982} but adapted to the needs of data analysis of gravitational wave signals as is shown in \cite{BSD1996}. The main focus is on presenting methods to deal with the inevitably occurring errors when subtracting the best-fit waveforms from the data stream. The residual errors have comparable spectral densities than the instrumental noise, because, roughly speaking, the subtraction of the best fit reduces the signal spectral density by a factor of $1/\rm SNR^2$. This is true for broad-band and narrow-band signals. Therefore, residual errors are too large to allow a measurement of an inflation generated background of gravitational waves. Fortunately, though arising from the presence of instrumental noise, the subtraction errors are not completely random but mostly confined to the tangent space of the template manifold at the point of the best fit. This restriction can be used to define a projection operator on the tangent space that cancels out all parts tangential to the manifold and hence most parts of the residual error \cite{CuHa2006}.

In section \ref{secMatched}, we give a short introduction to matched filtering and briefly point out the connection to differential geometry. Section \ref{secMaxLike} shows the derivation of the first-order approximation of the maximum likelihood estimator in case of high signal-to-noise ratio which provides important formulas and justifies the use of differential geometry. In the last section, we present the actual method of projection.

\subsection{Matched filtering and differential geometry}
\label{secMatched}
Most of the NS/NS signals observed by BBO will have amplitudes roughly two orders of magnitude smaller than the amplitude of the instrumental noise. Therefore some technique of filtering must be used to extract the information from the noisy data. Post-Newtonian expansions up to order 3.5 of the equations of motions of stellar objects, such as compact binary systems, yield very accurate waveforms throughout the BBO detection band, which can be employed to search for CBCs in the data streams. The fact that the shape of the signals is known to high accuracy is the reason why one can use \emph{matched filtering} which is also known as \emph{optimal filtering} since it provides the highest signal-to-noise ratio (SNR) of all linear filters \cite{WaZu1962}.

The detector outputs can, in two ways, be regarded as a vector. The first way which will be introduced in this section helps finding an expression for the SNR. In this case the detector output is a vector whose components are the outputs of the different detectors in different TDI channels.  For LISA-type detectors the output vector could be $\vec{\mathcal T}^{\rm s}(t) = (\mathcal T_A^{\rm s}(t), \mathcal T_E^{\rm s}(t), \mathcal T_T^{\rm s}(t))$ which is the one-detector case, for BBO the vector has 12 components, respectively. In case of a successful detection of a GW, the detector output is the sum of the signal $\mathcal T_i^{\rm h}(t)$ and additive, stationary, Gaussian noise $\mathcal T_i^{\rm n}(t)$, which are vectors in the same manner as described above. Optimal filtering of the data stream now means folding the output of the detector with a filter function $\vec{k}(t)$, and normalize it with the correlation of the instrumental noise. Henceforth, for ease of notation, we will drop the tilde ''$\tilde{\phantom{a}}$'' over frequency-domain functions and distinguish between time- and frequency domain functions by means of their arguments. Then, the multi-channel SNR can be defined in terms of scalar products between channel vectors:
\beq
    \SNR = \frac{\int \drm f \, \vec{k}(f) \cdot \vec{\mathcal T}^{\rm s}(f)}{\textnormal{rms} \int \drm f \, \vec{k}(f) \cdot \vec{\mathcal T}^{\rm n}(f)}
\eeq
The optimal filter function $\vec{k}(f)$ can be cast into the form
\beq 
    \vec{k}(f) = [\vec{\mathcal T}^{\rm m}(\vec\lambda,f)]^\dagger \, (S^{\rm n})^{-1}(f)
\eeq
where $\vec{\mathcal T}^{\rm m}(\vec\lambda,f)$ is the Fourier transform of the GW signal parameterized by a set of parameters denoted by $\vec\lambda$, the dagger means conjugate transpose. $S^{\rm n}(f)$ is the Fourier transform of the noise covariance matrix and its diagonal elements are the (double-sided) power spectral densities of the noise in the corresponding TDI channel. Hence it follows that
\beq
    \SNR(\vec\lambda) = \frac{\int \drm f \, [\vec{\mathcal T}^{\rm m}(\vec\lambda,f)]^{\dagger} \cdot (S^{\rm n})^{-1}(f) \cdot \vec{\mathcal T}^{\rm s}(f) }{\textnormal{rms} \int \drm f \, [\vec{\mathcal T}^{\rm m}(\vec\lambda,f)]^{\dagger} \cdot (S^{\rm n})^{-1}(f) \cdot \vec{\mathcal T}^{\rm n}(f) } \, ,
\eeq
$\vec{\mathcal T}^{\rm s}(f)$ the Fourier transform of the detector output, $\vec{\mathcal T}^{\rm n}(f)$ the noise in Fourier space. 

In the following we assume as in the previous sections that the noise of different detectors and channels is uncorrelated, and that we use the optimal TDI configuration. The optimal channels typically called A, E and T \cite{TEA2002b} are all statistically independent and hence, have uncorrelated noise contributions. In this case the correlation matrix of the noise is diagonal and with a new optimal filter function $k'_i(\vec\lambda,f) = [{\mathcal T}_i^{\rm m}(\vec\lambda,f)]^* / S^{\rm n}_i(f)$ the SNR can be rewritten in the form,
\beq
    \SNR(\vec\lambda) = \frac{ \sum\limits_i  \int \drm f \, k'_{i}(\vec\lambda, f) \, {\mathcal T}_i^{\rm s}(f) }{ \textnormal{rms} \sum\limits_{i} \int \drm f \, k'_i(\vec\lambda, f) \, {\mathcal T}_i^{\rm n}(f)}
\eeq
Here $S^{\rm n}_i(f)$ is the noise spectral density in the i\emph{th} optimal TDI channel. The index $i$ runs over all detectors and channels.

However, at this point it is advantageous to regard the data of each channel ${\mathcal T}_i^{\rm s}$ as a vector itself. Since a detector will sample data at a fixed frequency $f_{\rm s}$ ($\approx 10\,\textnormal{Hz}$ for BBO), each data stream will comprise $N = f_{\rm s} \cdot T$ measuring points, where $T$ is the total observation time (typically $10^8\,\textnormal{s}$ which is BBO's lifetime). Each measuring point at time $t_k = k / f_{\rm s}$ can be seen as a component of an $N$ dimensional vector ${\mathcal T}_i^{\rm s} = ({\mathcal T}_i^{\rm s}(t_1), {\mathcal T}_i^{\rm s}(t_2), \dots , {\mathcal T}_i^{\rm s}(T))$ in the vector space $\mathcal{V}_i$ of all detector outputs of channel $i$. In the first place, it is this definition of a data vector which underlies the geometrical interpretation presented in the following paragraphs, not necessarily the gathering of detector outputs into a channel vector. The outputs of all channels form a $12\,N$-dimensional vector space $\mathcal{V}$ which formally can be thought of as a direct sum of the $\mathcal{V}_i$. The instrumental noise ${\mathcal T}_i^{\rm n}$ and the gravitational wave signal ${\mathcal T}_i^{\rm h}$ are vectors in the same manner. Due to the fact that each binary signal is described by a set of 9 parameters (neglecting spin and eccentricity), the complete signal formed by about $10^4 - 10^5$ NS/NS with redshifts $z<8$ will be parameterized by $N_{\rm P} = 10^5 -10^6$ parameters and hence will lie on a submanifold $\mathcal{M}$ in $\mathcal{V}$ with dimension $N_{\rm P}$.

In case of stationary, Gaussian instrumental noise the matched filter induces an inner product on $\mathcal{V}$ that is defined by 
\beq
    \braket{\vec g}{\vec h} \equiv \sum\limits_{i} \int\limits_{0}^{\infty} \drm f \, \frac{g_i^*(f) \, h_i(f) + g_i(f) \, h_i^*(f)}{S^{\rm n}_i(f)}\,.
    \label{eqn:innerprod}
\eeq
It is straightforward to show that the ensemble average of $\braket{\vec g}{\vec{\mathcal T}^{\rm n}}\braket{\vec{\mathcal T}^{\rm n}}{\vec h}$ is equal to $\braket{\vec g}{\vec h}$ for an ensemble of realizations of instrumental noise, which can be proved by using $\overline{{\mathcal T}^{\rm n}(f){\mathcal T}^{\rm n}(f^\prime)} = \delta(f-f')S^{\rm n}(f)$ (see also \cite{CuHa2006,BSD1996}). In terms of this inner product, the optimal signal-to-noise ratio, with help of the above property, can easily be written as
\beq
    \SNR(\vec\lambda) = \braket{\vec{\mathcal T}^{\rm m}(\vec\lambda)}{\vec{\mathcal T}^{\rm m}(\vec\lambda)}^{\frac{1}{2}} 
    \label{eqn:optSNR}
\eeq
The inner product defined in equation~(\ref{eqn:innerprod}) has the same features than the scalar product in Euclidean vector spaces, i.e. it is positive-definite, and can therefore be used as a measure of distance and angles within the vector space. The length in Euclidean space corresponds to the total SNR Eq.~(\ref{eqn:optSNR}) collected by all channels. Angles quantify the correlation of two outputs, e.g. two outputs are orthogonal if their correlation vanishes. Thus the definition of an inner product enables one to establish a geometrical description of the problem of filtering.

In the high SNR limit, the best-fit parameters are found by maximizing the likelihood function, see Eq.~(\ref{eqn:likelihood}), which is equivalent to minimizing the inner product $\braket{\vec{\mathcal T}^{\rm s}-\vec{\mathcal T}^{\rm m}(\vec\lambda)}{\vec{\mathcal T}^{\rm s}-\vec{\mathcal T}^{\rm m}(\vec\lambda)}$, which can be considered as the distance of the detector output to the template manifold. Hence the best-fit template waveform is the one which has least separation to the detector output. In other words, geometrically, the best fit corresponds to the projection of the output $\vec{\mathcal T}^{\rm s}$ onto the submanifold $\mathcal M$, for more details see sections \ref{secMaxLike} and \ref{secProjection}.

To finish this section it should be said that generally the template manifold is not flat but rather has a curvature varying with the values of the parameters of the true signal. Like in general relativity this can be addressed by introducing a metric on the manifold. One possible choice for the metric is the covariance matrix of the parameter errors which also is the inverse of the Fisher information matrix (see section B), defined by
\beq
     \Gamma_{\alpha \beta} = \braket{ \frac{\partial \vec{\mathcal T}^{\rm m}(\vec\lambda)}{\partial \lambda^{\alpha}} }{\frac{\partial \vec{\mathcal T}^{\rm m}(\vec\lambda)}{\partial \lambda^{\beta}}}
\eeq
This statement will not be proven in this paper but the interested reader shall be referred to \cite{BSD1996}.\\

\subsection{Maximum likelihood estimator in the high signal-to-noise ratio limit}
\label{secMaxLike}
In this section we briefly derive the deviation of the best-fit parameters $\hat{\vec\lambda}$ from the true ones $\vec\lambda_0$ in the limit of high SNR, assuming that the signal model is accurate. The outcome will give the desired results in terms of the geometrical quantities introduced in the previous section. To obtain the results we use the Bayesian estimator and write the exponent of the posterior distribution as a Taylor series around the best-fit parameters. For more information and different ways of deriving the parameter errors see \cite{Val2007}.

First consider the Bayesian estimator of the parameter error that we name $\vec\epsilon = \vec\lambda - \vec\lambda_0$ and is defined by,
\beq
    \expec{\epsilon^{\alpha}} = \int \drm^{N_{\rm P}}\epsilon \, \epsilon^{\alpha} \, p(\vec\epsilon\,| \vec{\mathcal T}^{\rm s})
\eeq
The function $p(\vec\epsilon\,| \vec{\mathcal T}^{\rm s})$ is the \emph{posterior} distribution which determines the probability of a model determined by parameter-value deviations $\vec\epsilon$ from the true signal, given a measurement $\vec{\mathcal T}^{\rm s}$. The connection between posterior and likelihood function is the following
\beq
    p(\vec\epsilon\,|\vec{\mathcal T}^{\rm s}) = \mathcal{N} \, p_0(\vec\epsilon\,) \, p(\vec{\mathcal T}^{\rm s}|\vec\epsilon\,),
\eeq
where $\mathcal{N}$ is a normalization constant, $p_0(\vec\epsilon\,)$ comprises the \emph{prior} knowledge of parameter values and $p(\vec{\mathcal T}^{\rm s}\,|\vec\epsilon\,)$ is the likelihood of data $\vec{\mathcal T}^{\rm s}$ given a model $\vec\epsilon$. At the moment, only flat a priori probabilities are put into our simulation which means $p_0(\vec\epsilon\,) \equiv 1$ and the estimator can be written as
\beq
    \expec{\epsilon^{\alpha}} = \mathcal{N} \int \drm^{N_{\rm P}} \epsilon \, \epsilon^{\alpha} \, p(\vec{\mathcal T}^{\rm s}|\vec\epsilon\,),
\eeq
depending only on the likelihood function. In this case the normalization constant is the inverse of the integral over the likelihood over the whole parameter space.

As shown in \cite{Fin1992}, the likelihood can be expressed with help of Eq.~(\ref{eqn:innerprod}) as
\beq
    p(\vec{\mathcal T}^{\rm s}|\vec\lambda) \propto \exp\left( -\frac{1}{2} \braket{\vec{\mathcal T}^{\rm s} - \vec{\mathcal T}^{\rm m}(\vec\lambda)}{\vec{\mathcal T}^{\rm s} - \vec{\mathcal T}^{\rm m}(\vec\lambda)}\right) 
    \label{eqn:likelihood}
\eeq
Due to the fact that the data is a sum of instrumental noise and a signal, one can rewrite the argument within the brackets of Eq.~(\ref{eqn:likelihood}) as $\vec{\mathcal T}^{\rm n} + \delta\vec{\mathcal T}^{\rm m}$ where $\delta \vec{\mathcal T}^{\rm m} = \vec{\mathcal T}^{\rm m}(\vec\lambda_0) - \vec{\mathcal T}^{\rm m}(\vec\lambda) $. Here, $\vec{\mathcal T}^{\rm m}(\vec\lambda_0)$ represents the true signal $\vec{\mathcal T}^{\rm h}$ in terms of the accurate model. Now, the difference in the waveforms can be expanded in a Taylor series as
\beq
   - \delta \vec{\mathcal T}^{\rm m} = \partial_{\alpha} \vec{\mathcal T}^{\rm m}(\vec\lambda_0) \, \epsilon^{\alpha} + \frac{1}{2} \, \partial_{\alpha} \partial_{\beta} \vec{\mathcal T}^{\rm m}(\vec\lambda_0) \, \epsilon^{\alpha} \epsilon^{\beta} + \mathcal{O}(\epsilon^3)
    \label{eqn:deltaHtaylor}
\eeq
Inserting Eq.~(\ref{eqn:deltaHtaylor}) into Eq.~(\ref{eqn:likelihood}) and introducing normalized waveforms $\vec{\mathcal N}^{\rm m} \equiv \vec{\mathcal T}^{\rm m} / A$ and $\vec{\mathcal N}^{\rm m}_\alpha \equiv \partial_{\alpha} \vec{\mathcal T}^{\rm m} / A$ with $A = \braket{\vec{\mathcal T}^{\rm m}(\vec\lambda_0)}{\vec{\mathcal T}^{\rm m}(\vec\lambda_0)}^{1/2}$, which helps to highlight the dependence of the equations on the SNR, yields the following approximation of the likelihood:
\beq
\begin{split}
    p(\vec{\mathcal T}^{\rm s}|\vec\epsilon\,) \propto & \exp\bigg(-\frac{1}{2} \Big[ \braket{\vec{\mathcal T}^{\rm n}}{\vec{\mathcal T}^{\rm n}} -2 A \braket{\vec{\mathcal T}^{\rm n}}{\vec{\mathcal N}^{\rm m}_{\alpha}(\vec\lambda_0)}\,  \epsilon^{\alpha} \\
& + A^2 \left( \braket{ \vec{\mathcal N}^{\rm m}_{\alpha}}{ \vec{\mathcal N}^{\rm m}_{\beta}} - \frac{1}{A} \braket{\vec{\mathcal T}^{\rm n}}{\vec{\mathcal N}^{\rm m}_{\alpha \beta}} \right) \cdot \epsilon^{\alpha} \epsilon^{\beta} + \mathcal{O}(\epsilon^3) \Big]\bigg)
\end{split}
\eeq

The expansion can be cut off at order $\epsilon^3$ because higher order corrections in the exponent would correspond to higher order corrections to the estimator and we are only interested in the first order terms. One sees that for high SNR the likelihood is approximated by a multivariate normal distribution. The first summand is just a constant and will be absorbed into the normalization constant $\mathcal{N}$, the second term shifts away the maximum of the distribution from the true parameters due to instrumental noise, whereas the third term contains, in round brackets, the inverse of the covariance matrix of the errors and mainly determines the width of the distribution. The correlation of the noise with the second derivatives of the signal gives a first correction to the Fisher matrix as the inverse of the covariance but as will be clear from Eq.~(\ref{eqn:paramerrorAppr},) one can neglect this correction for high SNR since it scales with $1/A$ compared to the Fisher matrix.

With all that at hand one can compute the Bayesian estimator by solving the integral over the likelihood, which is a lengthy but straightforward calculation. Here we present just the result,
\beq
\begin{split}
\expec{\epsilon^{\alpha}} &= \frac{1}{A} \, \left[ \braket{ \vec{\mathcal N}^{\rm m}_{\alpha}}{ \vec{\mathcal N}^{\rm m}_{\beta}} - \frac{1}{A} \braket{\vec{\mathcal T}^{\rm n}}{\vec{\mathcal N}^{\rm m}_{\alpha \beta}} \right]^{-1} \braket{\vec{\mathcal T}^{\rm n}}{\vec{\mathcal N}^{\rm m}_{\beta}}  \\
                   &\approx \Gamma^{\alpha \beta} \braket{\vec{\mathcal T}^{\rm n}}{\vec{\mathcal T}^{\rm m}_{\beta}}
\end{split}
\label{eqn:paramerrorAppr}
\eeq
This equation reveals that, as promised, the deviations from the true parameters are completely determined by the template-manifold metric and the length of the projection of the noise vector onto the tangent space of $\mathcal{M}$ at the point of the best fit. Also, the parameter errors decrease with $1/\rm SNR$. We stress again that this result is obtained as a first-order $\SNR$ approximation which is supposed to be sufficient for BBO as the expected SNRs are high enough to justify this approach. Anyway, higher order expansions and the influence of \emph{prior} information on the estimator can be looked up in \cite{CuFl1994} and \cite{Val2007}. 

\subsection{The Projection Operator}
\label{secProjection}
In the last two sections, we outlined a strong connection between differential geometry and methods used in data analysis of gravitational waves from compact binary objects such as neutron-star neutron-star binaries. Investigations of matched filtering of data led to a definition of a scalar product on the vector space $\mathcal{V}$ of detector outputs. Modelling the waveforms by post-Newtonian templates smoothly depending on a set of parameters describing the physical properties of the binaries, the detector as well as their relative motion, confines the possible outputs generated by a gravitational wave to a submanifold $\mathcal{M}$ within $\mathcal{V}$. The errors occurring at the estimation of the signal parameters are then completely given in terms of geometrical quantities such as projections onto and within the tangent space of $\mathcal{M}$ at the best-fit parameters.

In this section, we further exploit the geometrical restrictions on the waveforms and present a successful implementation of a projection method to cancel the subtraction noise $\delta \vec{\mathcal T}^{\rm m}$ occurring by subtracting the best fit from the data stream. Eq.~(\ref{eqn:paramerrorAppr}) showed that the expected parameter errors are proportional to $1/\SNR$ which can be used to see how the amplitude of the subtraction noise depends on the SNR. The following equation provides a Taylor expansion of the template waveform around the true signal $\vec{\mathcal T}^{\rm m}(\vec\lambda_0)$ evaluated at the expected error:
\beq
\begin{split}
    \vec{\mathcal T}^{\rm m}(\expec{\vec\epsilon\,}) &= \vec{\mathcal T}^{\rm m}(\vec\lambda_0) + \vec{\mathcal T}_{\alpha}^{\rm m}(\vec\lambda_0)\expec{\epsilon^{\alpha}} + \vec{\mathcal T}_{\alpha\beta}^{\rm m}(\vec\lambda_0) \expec{\epsilon^{\alpha}}\expec{\epsilon^{\beta}}+ \mathcal{O}(\SNR^{-2})
\end{split}
    \label{eqn:Htaylor}
\eeq
Making use of the fact that $\vec{\mathcal T}^{\rm m}(\vec\lambda_0)$ and all derivatives of $\vec{\mathcal T}^{\rm m}$ depend linearly on the SNR, one finds that the first term in this expansion is proportional to SNR, the second which is the leading term of the subtraction noise is independent of the SNR which can be seen, too, by computing the mean norm of the subtraction noise to leading order in the SNR
\beq
    \begin{split}
    \overline{ \braket{\delta \vec{\mathcal T}^{\rm m}(\vec\epsilon\,)}{\delta \vec{\mathcal T}^{\rm m}(\vec\epsilon\,)} } &= \braket{\vec{\mathcal T}^{\rm m}_{\alpha}(\vec\lambda_0)}{\vec{\mathcal T}^{\rm m}_{\beta}(\vec\lambda_0)} \overline{ \epsilon^{\alpha} \epsilon^{\beta}} \\
                                         &= N_{\rm P}
    \end{split}
\label{eqSubNoise}
\eeq
Again, $N_{\rm P}$ is the number of parameters describing the total signal, or in other words the dimension of the template manifold. The mean spectral density of the subtraction noise in each channel is most suitably expressed in terms of the ratio $S(\delta{\mathcal T}_i^{\rm m};\; f) / S^{\rm n}(f)$ of spectral densities of the subtraction and the instrumental noise. In the simplest possible case, we could consider a signal with constant ratio $S({\delta{\mathcal T}_i^{\rm m}};\; f) / S^{\rm n}(f)$ for each frequency within a given signal bandwidth and negligible ratios outside the bandwidth. The integral which yields the scalar product in Eq.~(\ref{eqSubNoise}) is converted into a sum of these ratios over all frequency bins at $f = 1/T, \dots, f_{\rm s} / 2$ within the signal bandwidth. For BBO which furnishes data from 8 channels sensitive to GWs and which permits a total observation time of $T=10^8\,$s, and assuming a total signal of $10^5$ CBCs ($N_{\rm P}\sim 10^6$) is contained within a bandwidth of 1\,Hz, this would lead to $S({\delta{\mathcal T}^{\rm m}_i};\; f )/ S^{\rm n}(f) = 0.5 \, N_{\rm P} / (8\cdot 1\,{\rm Hz} \cdot T) \sim 10^{-3}$. A cosmic gravitational wave background with energy density $\Omega \gtrsim 10^{-17}$ would have a spectral density which is more than one order of magnitude less than the weakest possible subtraction noise level. This level is based on an optimal search of the CBCs which cannot be performed even if a steady development of computational facilities over two or three decades in accord with Moore's law is assumed. So the subtraction noise, if remaining within the data, would prohibit the detection of an inflation generated background with high certainty.

Eq.~(\ref{eqn:Htaylor}) also shows that the third term is proportional to $1/\SNR$. So in deleting the zeroth and first order terms from the data stream one reduces the signal strength by a factor of $1/\SNR^2$. The first order term is a linear combination of first derivatives of the signal, or in other words a vector lying in the tangent space of $\mathcal{M}$ at the true parameter values. Since the expected parameter errors scale with 1/SNR, it follows that the two tangent spaces at the best-fit and the true signal can be regarded as nearly identical. So from now on, all derivatives are taken at the best-fit parameters which are obtained by searching the data for signals and estimating model parameters. The leading term of the subtraction noise is taken as a vector in the tangent space at the best fit.

\begin{figure}[htbp]
\hspace*{-0.5cm}\includegraphics[width=9cm, height=6cm]{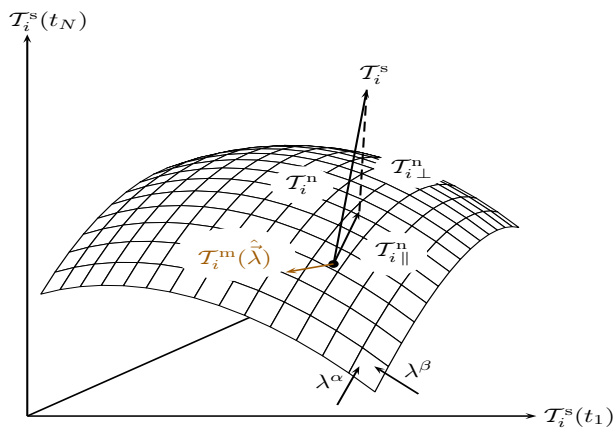}
\caption{(Color online) The figure represents the template manifold $\mathcal{M}_i$ of a specific TDI channel $i$. Since the best-fit parameters $\hat{\vec\lambda}$ are estimated with information from all channels, the projection of data ${\mathcal T}_i^{\rm s}$ in channel $i$ does not yield the best-fit waveform ${\mathcal T}_i^{\rm m}(\hat{\vec\lambda})$. It is assumed that the true signal lies on the template manifold at the marked point. The noise vector ${\mathcal T}_i^{\rm n}$ which points from the true signal to the measured data, is split into its components parallel and perpendicular to the manifold at the true signal. The (tiny) vector of the CGWB is not shown.}
\label{fig:SubDataSpace}
\end{figure}

Fig.~(\ref{fig:SubDataSpace}) schematically shows the template manifold $\mathcal{M}_i$ of signals contained in the data stream of channel $i$, the true signal ${\mathcal T}_i^{\rm h}$, the best-fit ${\mathcal T}_i^{\rm m}(\hat{\vec\lambda})$ and the parallel and perpendicular parts of the instrumental noise. In a specific TDI channel, the projection of instrumental noise generally does not coincide with the difference between the best-fit template and the true signal, as it is the case in the complete vector space $\mathcal{V}$. That is due to the fact that by using information from all channels to find the best-fit, your estimate will be somewhat better than if you had determined the parameters just with one data stream. Important to say is that because the best fit subtracted from the data in each channel is determined by the same best-fit parameters, the subtraction noise will be correlated in all channels and not be deleted by a cross-correlation measurement between channels, see section \ref{secCross}. But good news is that most of the correlated error is restricted to the tangent space of the manifolds $\mathcal{M}_i$ at the best fit, and projection out tangential directions of the residual data will delete the correlations. The tangential part of the CGWB which gets projected out is negligible \cite{CuHa2006}.

To perform this task one can define a projection operating on a specific channel $i$, which removes the tangential parts:
\beq
    \mathsf{P}_i = \mathbbm{1} - \Gamma_i^{\alpha \beta} \, |\partial_\alpha{\mathcal T}_i^{\rm m} \rangle \langle \partial_\beta{\mathcal T}_i^{\rm m}|
    \label{eqProjOP}
\eeq
where $\Gamma_i^{\alpha \beta}$ denotes the inverse of the Fisher matrix $\Gamma^i_{\alpha\beta}=\braket{\partial_\alpha {\mathcal T}_i^{\rm m}(\hat{\vec\lambda})}{\partial_\beta {\mathcal T}_i^{\rm m}(\hat{\vec\lambda})}$. The projected data stream can be calculated, e.g.~in the frequency domain, according to
\beq
    \mathsf{P}_i{\mathcal T}^{\rm res}_i(f) = {\mathcal T}_i^{\rm res}(f) - \Gamma_i^{\alpha \beta} \,\braket{ \partial_\beta{\mathcal T}_i^{\rm m}}{{\mathcal T}_i ^{\rm res}}\partial_\alpha{\mathcal T}_i^{\rm m}(f)
\eeq
Here, ${\mathcal T}^{\rm res}_i$ denotes the residual data which remains after subtraction of the best fits. In case the initial data stream contains a GW foreground ${\mathcal T}_i^{\rm h}$ generated by CBCs, a CGWB ${\mathcal T}_i^{\rm b}$ and instrumental noise ${\mathcal T}_i^{\rm n}$, then after subtracting the best fit and projecting the data, what is left is
\beq
    \mathsf{P}({\mathcal T}^{\rm res}_i) = {\mathcal T}^{\rm n}_{i\,\perp} + {\mathcal T}^{\rm b}_{i\,\perp} + \mathcal{O}({\mathcal T}_i^{\rm h}/\SNR^{2})
\eeq
the perpendicular parts of the instrumental noise and the cosmic GW background. The loss of power in ${\mathcal T}^{\rm b}_i$ is negligible as already mentioned and the remnants of the instrumental noise will be removed by cross-correlating the data of the two collocated detectors in the star-of-David configuration.

We want to point out another interesting property of the subtraction-noise projection. For arbitrary template manifolds, the best fit has to lie close to the true signal. If that is not the case, then tangential planes at the best fit and the true signal do not coincide. This could entail a poor performance of the projection operator Eq.~(\ref{eqProjOP}) in removing residual power of the signal spectrum. Now imagine a flat template manifold. In that case, the best fit does not have to be accurate, since tangential planes coincide at all points on the manifold. In fact, since the vanishing signal is always an element of the model (e.g.~coming from a source which is far away), the projection method works without having to estimate the signals at all! The only information which is needed is the number of signals. This information determines the right dimension of the manifold. Although a completely flat metric does not represent most analysis problems, we can already see in (signal $\rightarrow$ subtraction noise $\rightarrow$ projected signal) spectra shown in section \ref{secResProj}, that the power of the projected spectrum is very low at all frequencies independent of the power of the subtraction noise. Obviously in our simulation, some parameters of some signals were poorly estimated, but their subtraction noise is removed with high accuracy. This can just be explained with a sort of ''partial'' flatness of the template manifold. Certainly, this feature needs to be investigated in the future.

\section{Cross correlation of TDI channels}
\label{secCross}
In section \ref{secORF} we have seen, that the CGWB is completely covered by the detector noise. One needs to reduce the difference between noise and background spectral densities by a factor of $10^2$ in the case of a CGWB with $\Omega=10^{-15}$. In this section we will describe, how that reduction can be done by cross correlating the TDI streams of the collocated detectors. Increasing the SNR of the background is achieved by increasing the observation time until the correlation output is dominated by contributions from the CGWB. One can study the correlation measurement for small observation times $T$ and then extrapolate the output for higher $T$ by making use of a simple scaling law of the SNR with observation time. We will see that performing a correlation measurement with data gathered over 3 years, which is BBO's proposed mission lifetime, it is possible to detect a CGWB with energy densities below $\Omega=10^{-16}$. These results are based on the assumption that the subtraction noise from the CBC foreground can be removed with sufficient accuracy. Our results which are presented in section \ref{secResults} show that this is indeed the case (at least for 100 NS/NS).

We consider the TDI channel output ${\mathcal T}^{\rm s}_i(f)$ as a sum of the CGWB and of the detector noise,
\begin{equation}
{\mathcal T}^{\rm s}_i(f) = {\mathcal T}^{\rm b}_i(f)+ {\mathcal T}^{\rm n}_i(f),
\end{equation}
where the noise is assumed to be Gaussian, stationary and uncorrelated between different channels, and the foreground signal is subtracted below the CGWB. Unlike the instrumental noise, the CGWB will be correlated in different channels to some degree which can be predicted by the ORF. In the frequency domain, the expected outcome of a correlation $C_{ij}$ between channels $i,j$ is an integral over all frequencies of the data-stream product
\begin{equation}
\begin{split}
C_{ij} & = \int \drm f \, {\rm Re}\left\{{\mathcal T}^{\rm s}_i(f)[{\mathcal T}^{\rm s}_j(f)]^*\right\}Q_{ij}(f)\\
& = \sum\limits_f \frac{{\rm Re}\left\{{\mathcal T}^{\rm s}_i(f)[{\mathcal T}^{\rm s}_j(f)]^*\right\}}{T}Q_{ij}(f)
\end{split}
\label{eqCorrelate}
\end{equation}
where a channel-dependent filter function $Q_{ij}$ is used to suppress contributions from frequencies with strong instrumental noise or weak (expected) CGWB. According to Eq.~(\ref{eqCGWBCorr}), the expectation value $\expec{C_{ij}}$ of the correlation measurement for sufficiently long observation times is
\begin{equation}
 \expec{C_{ij}} = \sum\limits_f \gamma_{ij}(f)S^{\rm b}(f)Q_{ij}(f)
 \label{eqn:CorrExpec}
\end{equation}
where $S^{\rm b}$ is the GW strain spectral density. The variance of correlation noise which is dominated by contributions from the instrumental noise is given by
\begin{equation}
\expec{(\Delta C_{ij})^2} = \sum\limits_f S^{\rm n}_i(f)S^{\rm n}_j(f)\,Q_{ij}^2(f)
 \label{eqn:CorrDev}
\end{equation}
Now, we can understand how in general the correlation signal-to-noise ratio $\SNR_{ij} = \expec{C_{ij}}/\sqrt{\expec{(\Delta C_{ij})^2}}$ scales with observation time $T$. The number of frequencies (frequency bins) summed over in Eq.~(\ref{eqn:CorrExpec}) increases proportional to $T$ whereas the standard deviation, which is the square root of Eq.~(\ref{eqn:CorrDev}), scales with $\sqrt{T}$. Therefore, increasing the observation time, one eventually raises contributions from the CGWB above the expected deviations. If the range of frequencies $\Delta f$ contained in the sums is small enough, then the functions within the summands can be taken as constants and the two equations, evaluated at a fiducial frequency $f_0$ which lies within the bandwidth, become
\beq
\begin{split}
\expec{C_{ij}} & \approx (T\cdot\Delta f)\gamma_{ij}(f_0)S^{\rm b}(f_0)Q_{ij}(f_0)\\
\sqrt{\expec{(\Delta C_{ij})^2}} & \approx \sqrt{T\cdot\Delta f}\sqrt{S^{\rm n}_i(f_0)S^{\rm n}_j(f_0)}\,Q_{ij}(f_0)
\end{split}
\eeq
So, in this small-bandwidth approximation, the SNR is independent of the (constant) filter function $Q_{ij}$
\beq
\SNR_{ij} = \sqrt{T\cdot\Delta f}\frac{\gamma_{ij}(f_0)S^{\rm b}(f_0)}{\sqrt{S^{\rm n}_i(f_0)S^{\rm n}_j(f_0)}}
\eeq
If the small-bandwidth, flat spectrum approximation does not hold, then there exists an optimal filter which is based on models for the noise spectral densities in the two channels, the ORF and the spectral density of the stochastic background. Its purpose is to suppress contributions from frequencies which would contribute strongly to the instrumental noise in $C_{ij}$, but weakly to the GW correlations. Accordingly, the optimal filter is given by \cite{All1996}
\begin{equation}
 Q_{ij}(f) = \frac{\gamma_{ij}(f)S^{\rm b}(f)}{S_i^{\rm n}(f)S_j^{\rm n}(f)},
\label{eqOptFilter}
\end{equation}
\begin{figure}[ht]
\includegraphics[width=6cm,angle=-90]{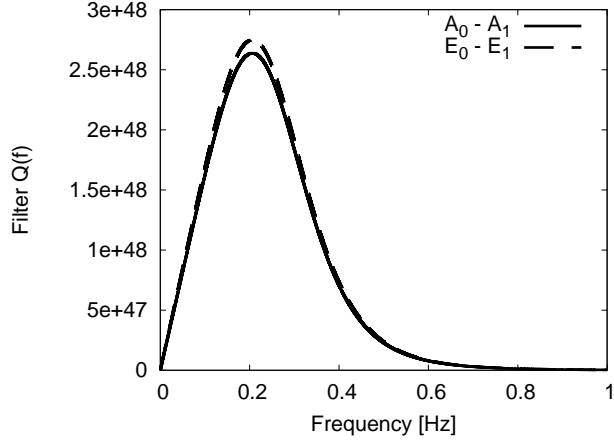}
\caption{The graph displays the filter functions $Q_{A_0A_1}(f)$ and $Q_{E_0E_1}(f)$. It contains models for the instrumental noise spectral densities (given in Eq.~(\ref{eqNoise})) and is based on a flat-$\Omega$ model of the CGWB.}
\label{figFilter}
\end{figure}
The filter function between channels $A_0$ and $A_1$ based on a flat $\Omega=10^{-15}$ model for the CGWB is shown in Fig.~\ref{figFilter}. In addition, one should keep in mind that the WD/WD barrier enforces a lower boundary on correlation frequencies. The filter maximum lies at $0.2\,$Hz which is half-way inside the WD/WD spectrum. In fact, most models presented in \cite{FaPh2003} predict a cosmological distribution of WD/WD which gives rise to an energy density of $\Omega\sim 10^{-14}$ -- $10^{-13}$ at 0.2\,Hz, which would make a detection of a cosmological background impossible at these frequencies unless the WD/WD signals were resolvable. Assuming that the WD/WD foreground at frequencies 0.2\,Hz cannot be analyzed, one has to find out if the maximum of the filter does determine the most efficient correlation frequencies. This is not the case, but as we will see, efficient frequencies are not much greater than suggested by the filter. Optimally, one had to design the instrument such that the most efficient frequencies lie above the WD/WD barrier. To find a definite answer to this problem, one has to calculate the contribution from certain frequencies to the SNR expected from a specific model of the CGWB. The SNR for the optimal filter assumes the form
\beq
\SNR_{ij}=\left(\sum\limits_f\frac{\gamma^2_{ij}(f)[S^{\rm b}(f)]^2}{S^{\rm n}_i(f)S^{\rm n}_j(f)}\right)^{1/2}
\label{eqSNRcorr}
\eeq
For a network of detectors, one would sum over all independent correlation pairs $(ij)$ to obtain the total network SNR
\beq
\begin{split}
\SNR_{\rm tot} & = \left(\sum\limits_{(ij)}\SNR^2_{ij}\right)^{1/2}\\
& = \left(\sum\limits_f\SNR_{\rm tot}^2(f)\right)^{1/2}\\
& = \left(\sum\limits_f\sum_{(ij)}\frac{\gamma^2_{ij}(f)[S^{\rm b}(f)]^2}{S^{\rm n}_i(f)S^{\rm n}_j(f)}\right)^{1/2}
\end{split}
\label{eqSNRnet}
\eeq
\begin{figure}[ht]
\includegraphics[width=6cm,angle=-90]{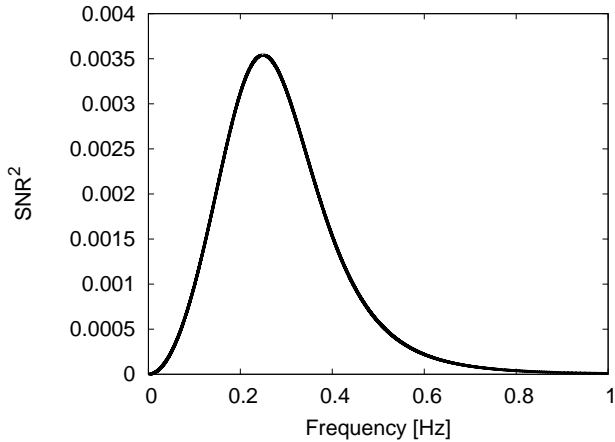}
\caption{The curve shows $\SNR^2_{\rm tot}(f)$ as defined in Eq.~(\ref{eqSNRnet}). It includes contributions from the two statistically independent pairs $A_0\leftrightarrow A_1$ and $E_0\leftrightarrow E_1$. This curve does not depend on the observation time $T$. The total $\SNR^2$ can be calculated by adding values of $\SNR^2_{\rm tot}(f)$ for each frequency bin $f_i=i/T$ within a chosen correlation bandwidth (a lower boundary for this band is set by the WD/WD barrier, the upper boundary is ultimately set by half of the sampling frequency $f_{\rm s}$). For example, neglecting the WD/WD barrier and approximating the area under the curve by a rectangle $(0.3\,{\rm Hz}) \times (0.003)$, a $\SNR^2=25$ would be obtained after $T=25/(0.3\cdot 0.003)\,{\rm s}\approx 3\cdot 10^4\,$s.}
\label{figSNRB}
\end{figure}
Fig.~\ref{figSNRB} shows $\SNR_{\rm tot}^2(f)$ including the two correlation pairs $A_0\leftrightarrow A_1$, $E_0\leftrightarrow E_1$ of BBO. The maximum of this curve is shifted towards higher frequencies with respect to the filter maximum, because the additional $\gamma_{ij}$ brings in a factor $f^4$ at frequencies below 1\,Hz and the additional background spectrum a factor $f^{-3}$. So, in total, the filter spectrum is multiplied by a factor $f$. Most of the SNR is collected at frequencies near 0.23\,Hz which may still be a bit too low. Certainly, this issue needs to be investigated in the future.

\section{Results}
\label{secResults}
Our results are presented in two ways. In the first part of this section, we show subtraction noise and projected signal spectra and compare them in total power. In the second part, the outcome of correlation measurements between the two collocated detectors is summarized in tables, and we compare contributions from instrumental noise, CGWB and CBCs. Also, the decrease of CBC correlations by subtraction of best fits and noise projection is investigated. A sensitivity of BBO to stochastic backgrounds is derived and extrapolated to an observation time of 3 years.

\subsection{Projection}
\label{secResProj}
In this section, we focus on two salient features of the projection results, which are predicted by theory. First, the projected spectra are compared with subtraction-noise spectra to find that the residual signal power is sufficiently small to enable CGWB detection within a certain frequency range. Second, we show that the projection operates selectively on subtraction noise and leaves stochastic signals like the instrumental noise and the CGWB more or less unaffected. 

\begin{figure}[ht]
\includegraphics[width=6.4cm,angle=-90]{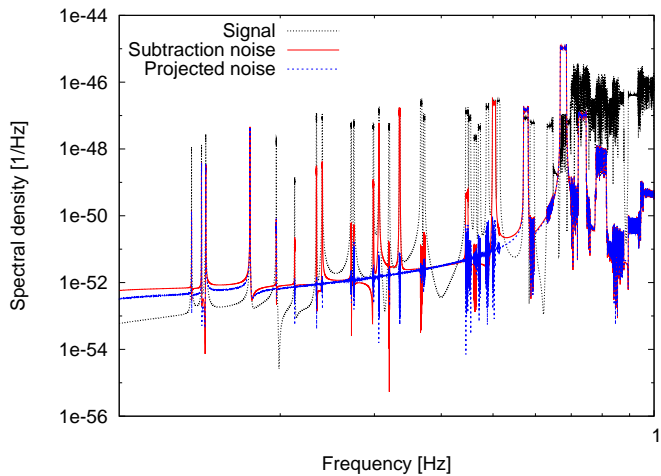}
\caption{The figure shows the recorded signal spectrum (dotted), the subtraction-noise spectrum (dashed) and the projected subtraction-noise spectrum (solid) between 0.1\,Hz and 1\,Hz. The subtraction noise is comparatively high at low frequencies, since pN waveforms are determined by strongly correlated parameters. These correlations decay at higher frequencies where a considerable part of the phase evolution is observed leading to better waveform estimates. Subtraction noise within the correlation band 0.2\,Hz --- 0.5\,Hz is projected. As one can see, all peaks of the subtraction noise are removed. The residual noise is negligible compared with a CGWB spectrum with fractional energy density $\Omega=10^{-15}$.}
\label{figProjSignal}
\end{figure}

In Fig.~\ref{figProjSignal}, the CBC spectrum together with the subtraction noise and its projection are displayed from 0.1\,Hz to 1\,Hz. With a few exceptions, the subtraction noise is weaker than the signal spectrum. The projection is applied to 17 out of 100 signals to remove the subtraction noise between 0.2\,Hz --- 0.5\,Hz. 
The peaks in the subtration-noise spectrum are removed. The residual spectrum lies below a CGWB with $\Omega=10^{-15}$ (see Fig.~\ref{figPrimordial}). Remarkably, the projection works accurately although some of the CBCs in the correlation band are poorly estimated, which can be seen by comparing the subtraction noise with the signal spectrum: the weaker the subtraction noise, the better the best fit. Simulating longer observation times would significantly improve the gain of the noise projection. We conclude that the subtraction noise will pose no problem to future-generation detectors as long as all foreground signals can be detected and modelled accurately.

Projecting the data means to remove all contributions which are correlated with certain template derivatives. These derivatives are associated with directions in a sampling space. A stochastic process has the property to distribute its energy randomly along all directions of the sampling space with a given mean value of the energy per direction, which is the geometrical interpretation of the fact that a stationary process has constant variance.
\begin{figure}[ht]
\includegraphics[width=6.4cm,angle=-90]{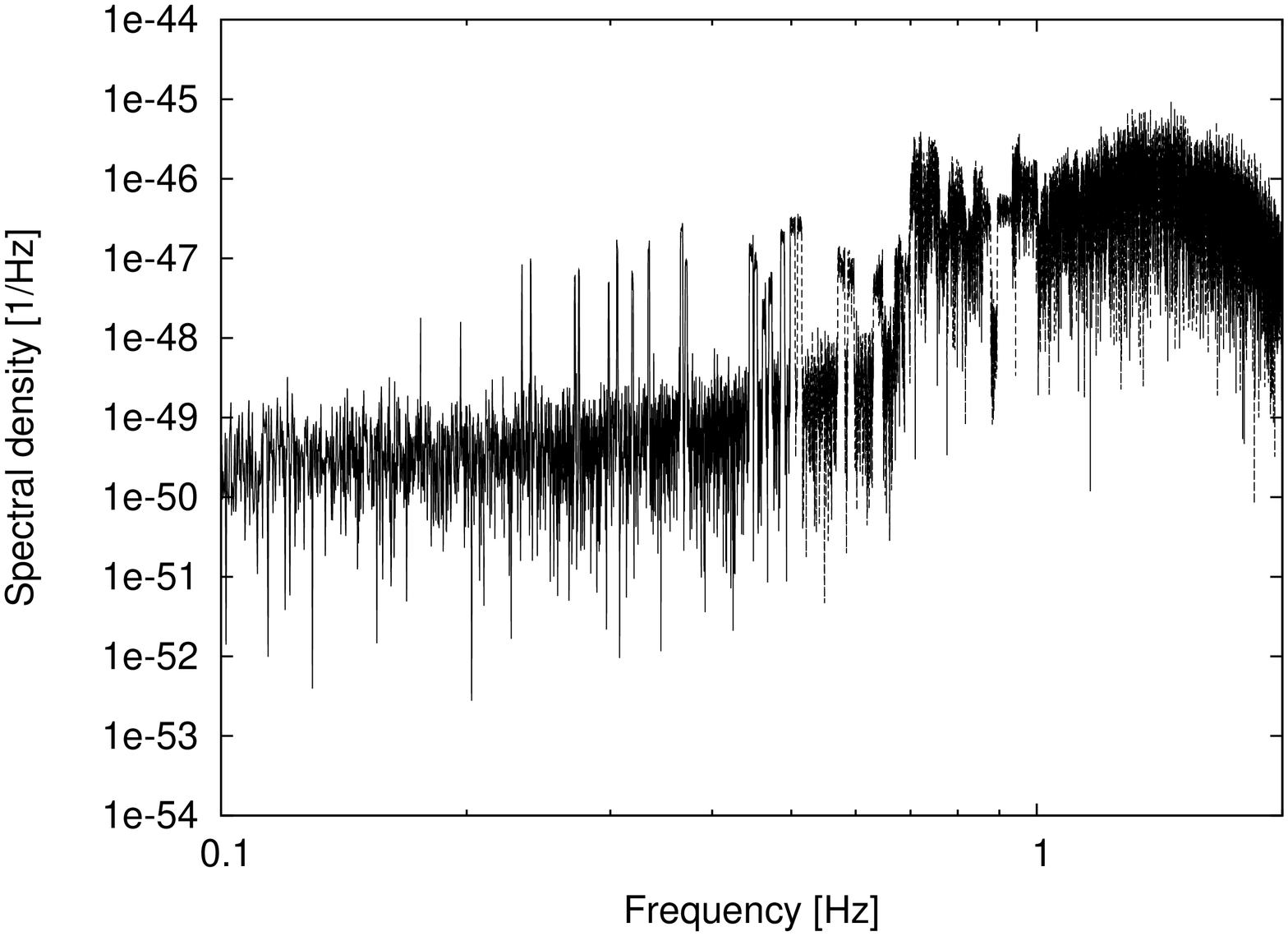}\\
\includegraphics[width=6.4cm,angle=-90]{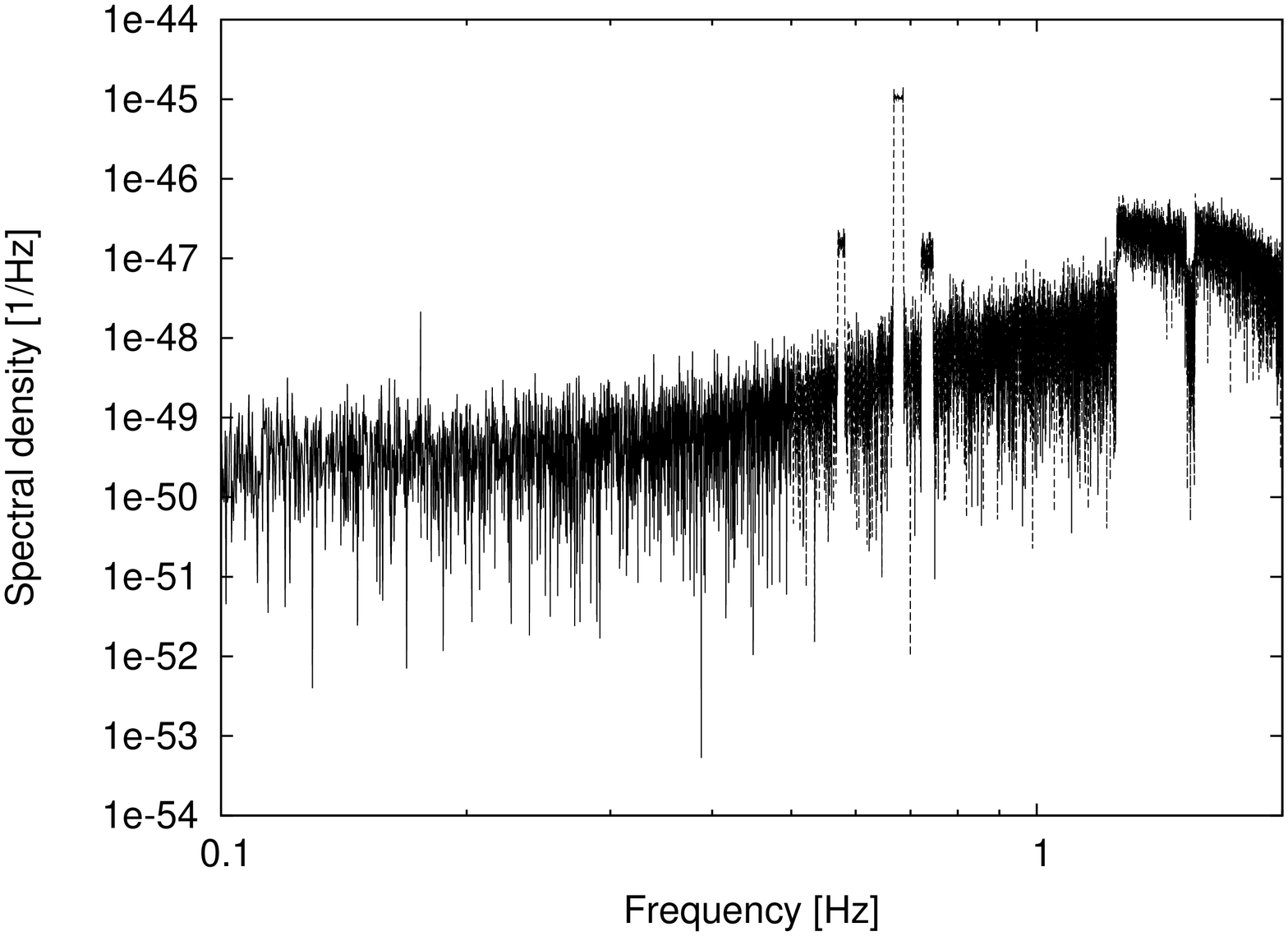}
\caption{The upper graph displays the spectrum of the total data before it is analyzed, whereas the lower graph displays the spectrum after best-fit subtraction and noise projection. Some of the signal peaks in Fig.~\ref{figProjSignal} can be identified in these two spectra. There is no visible change of the instrumental noise level. Here, every 20th frequency bin is plotted to reduce the amount of data.}
\label{figProjData}
\end{figure}
If we project out $N_{\rm p}$ directions of the sampling space, then, in average, we will remove a fraction $N_{\rm p}/N$ of the power of any stochastic, stationary process, where $N$ is the dimension of the sampling space which is the number of samples. As long as the number of projected directions is much smaller than $N$, the loss in power will be negligible. In our case, we project along $17\times 9$ (17 CBCs with 9 parameters each) out of $10^5\times 5.24288$ (observation time multiplied with sampling frequency) directions. The mean predicted power loss of stationary processes like the CGWB or the instrumental noise is 0.03$\%$ which is indeed insignificant. In Fig.~\ref{figProjData}, we plot the spectrum of the total data before (upper figure) and after (lower figure) data analysis including the noise projection. There is no visible change in the instrumental noise spectrum, but all signal peaks between 0.2\,Hz and 0.5\,Hz are removed. More quantitative results can be found in the next section which presents correlation values of all components of the total data.

\subsection{Correlation}
\label{secCorrelate}
The correlation measurement is evaluated in two ways. At first, we compute statistics for different observation times to make it possible to extrapolate our results to a 3 year observation time with certain confidence. These statistics are based on simulations without the CBC foreground. The reason is that the computation would take too long and also, the projection performance is insufficient at observation times much shorter than $T=10^5\,$s, because the Fisher matrices are highly ill-conditioned and parameter correlations are too strong to allow a meaningful estimation of signal parameters by means of Eq.~(\ref{eqParError}). Second, we investigate the correlation including the CBCs and noise projection for an observation time $T=10^5\,$s.

As mentioned in earlier sections, the correlation measurements are carried out between channels $A_0\leftrightarrow A_1$ and $E_0\leftrightarrow E_1$ and subsequently, the two values are added to form the total correlation SNR. Each correlation value corresponds to an integral over the correlation band 0.2\,Hz---0.5\,Hz. The results are shown in Table \ref{tabCorrelate}. 
\begin{table}
\begin{tabular}{|c||c|c|c|}
\hline
$T$ & $C_{\rm n}$ & $C_{\rm cgwb}$ & $C_{\rm tot}$ \\
\hline\hline
$1\cdot 10^{4} s$ &$0.00\pm 2.60$  & $5.90\pm 0.07$ &$5.90\pm 2.54$ \\
$2\cdot 10^{4} s$ &$0.83\pm 5.74$  & $14.72\pm 0.12$ &$15.34\pm 5.87$\\
$4\cdot 10^{4} s$ &$-0.93\pm 6.54$  & $26.28\pm 0.19$ &$25.56\pm 6.42$ \\
$8\cdot 10^{4} s$ &$1.73\pm 9.87$  & $53.80\pm 0.29$ &$55.20\pm 10.12$\\
\hline
\end{tabular}
\caption{The table contains the mean values and standard deviations of correlation measurements between channels $A_0\leftrightarrow A_1$ and $E_0\leftrightarrow E_1$ for different observation times. Values in each row are based on 20 measurements. Here, the total data is the sum of the CGWB and the instrumental noise. Whereas the mean value of the correlation $C_{\rm tot}$ of the total data is dominated by contributions from the CGWB, its standard deviation is determined by contributions from the instrumental noise. The mean value of $C_{\rm tot}$ is supposed to increase linearly with $T$, whereas its standard deviation -- predicted to be the square root of the mean value -- is supposed to increase with $\sqrt{T}$.}
\label{tabCorrelate}
\end{table}
The first column contains values of four different observation times used to obtain the correlation statistics. The second column shows the respective correlation outcome of the instrumental noise, the third column of the CGWB with $\Omega=10^{-15}$ and the last column of the total data, which, in this case, is the sum of the instrumental noise and the CGWB. Each correlation value is based on 20 measurements. In summary, the evolution of the mean values and standard deviations of the correlation with observation time confirms theoretical predictions. The mean value of $C_{\rm tot}$ increases approximately linearly with observation time $T$ and its standard deviation increases with the square root of $T$. Furthermore, the standard deviations of $C_{\rm tot}$ are always close to the square root of the mean value which would ideally hold, since the optimal filter defined in Eq.~(\ref{eqOptFilter}) is applied. The fact that the measured standard deviation is somewhat higher than the predicted value means, that our noise model which governs the correlation filter is weaker than the actually measured spectrum of the instrumental noise.  This descripancy can be explained since the (equal-arm) noise model is not used to generate time series of the noise in our simulation. In reality, one would take greater care in choosing the right noise model such that variances are equal to mean values, and the SNR is faithfully calculated for a single measurement. Alternatively, one may use the total spectrum after projection as noise model. In our simulation, we obtain the SNR by averaging over many measurements and the quality of the noise model does not have to be very high.

Before extrapolating to higher observation times, we have to make sure that subtraction noise can be removed from the data. Therefore, we show correlation values for our longest run with $T=10^5\,$s related to the CBC:
\beq
C_{\rm cbc}=19000,\quad C_{\rm sub}=3560,\quad C_{\rm proj}=2.34
\eeq 
The correlation value of the CBC without best-fit subtraction and projection is $C_{\rm cbc}$. When the best fits are subtracted, this value is reduced to $C_{\rm sub}$. By consequence, a CGWB with $\Omega=10^{-15}$ could not be detected without noise projection, since the CGWB correlation is around 55. The correlation of the projected noise is further reduced to $C_{\rm proj}$ which lies well below the CGWB value. The background becomes detectable. A remaining problem is that the true data will contain about 100 to 1000 times more signals which leads to a similar increase of the projected spectrum, but two effects are working for the good of the mission. Theoretically, the projected spectrum should decrease with $1/T$ which is due to an improvement of the accuracy of the estimated signal parameters. In addition, we believe that our projection results would be much better, if a different template class was used or if we had implemented the F statistics to maximize over nuisance parameters, since our waveforms are governed by templates with highly correlated parameters. These issues will have to be scrutinized in the future.

Now, assuming that the foreground subtraction noise can be projected out, we extrapolate correlation values to an observation time $T=10^8\,$s. Our measurments confirm that the SNR scales with the square root of the observation time, and so we extrapolate our results of Tab.~\ref{tabCorrelate} accordingly for each observation time and average over the four different predictions to obtain BBO's SNR corresponding to a flat stochastic background (i.e.~constant $\Omega$):
\beq
\SNR(3\,{\rm years}) \approx 200\cdot\frac{\Omega}{10^{-15}}
\eeq
One has to keep in mind that this prediction is based on a restricted correlation band 0.2\,Hz---0.5\,Hz. The SNR would double if the entire detection band especially towards lower frequencies was chosen as correlation band. The lower boundary is understood as cut-off frequency due to the WD/WD foreground which may not be analyzable below 0.2\,Hz. The upper boundary has no significant effect since high-frequency contributions to the SNR of the cosmic background are negligible. If one demands a minimal SNR of 5, then BBO's sensitivity is
\beq
\Omega_{\min}\approx 2.5\cdot 10^{-17}
\eeq
Given a lower bound of 0.2\,Hz for correlation measurements, there should exist an optimal arm length of BBO which is slightly smaller than 50000\,km, and which leads to a slightly improved sensitivity. It may also turn out that our choice for the lower boundary was too pessimistic and that BBO is more sensitive even without modifications.

\section{Conclusion}
The analysis presented in this paper serves three main purposes. The primary intention is to demonstrate the application of a subtraction-noise projection on a simulated data-anaylsis problem. The method is explained by invoking a geometrical interpretation of optimal signal detection and parameter estimation in the presence of additive, Gaussian noise. Second, we showed how to construct an analysis pipeline for a time-delay interferometer network, which seeks for a stochastic gravitational-wave background in the presence of a foreground built of compact-binary coalescences (CBCs). The detector network is simulated dynamically, thereby automatically including detector motion and time-varying response functions. Therefore, the definition of the overlap-reduction function had to be generalized so that it directly quantifies the total instrumental influence on correlation strengths of detector outputs, instead of quantifying the correlation of projected gravitational-wave induced strains in terms of somehow normalized quasi-stationary detector response functions. Third, since our simulation is based on a design proposal for the BBO detector network, we derive a prediction of its sensitivity towards stochastic backgrounds, which should be more robust than values obtained from previous investigations.

The simulation creates time series of the instrumental noise and compact binaries for each photodiode in the network. During observation, response functions and detector positions change. We use equations of the orbital motion of each satellite which are specified up to second order in the orbital eccentricity. Consequently, each triangular configuration of satellites is simulated with cartwheel and breathing motion. The isotropic stochastic background with a given spectral density is directly generated in the frequency domain, by first computing transfer functions and overlap-reduction functions of and between each network output. These frequency-domain functions are obtained via FFT of the detectors' impulse responses. 

Based on the assumption that all foreground signals in the data from compact binaries are detected, we found as expected, that by removing the estimated foreground from the data, the residual foreground spectrum due to inaccurate fitting of waveforms covers the spectrum of the cosmological background and makes it impossible for BBO to detect it, unless its fractional energy density assumes very high values. We implemented a subtraction-noise projection into the pipeline, which allows to accurately remove the subtraction noise. In simulation runs with $T=10^5\,$s and operating on data with 100 CBCs, the total power of the projected signal is 4 orders of magnitude weaker than the original signal spectrum, which is much better than previously expected since the analysis is based on a comparatively short observation time which entails strong correlations between different parameters of the signal model. This outcome can just be explained by assuming that the template manifold is approximately flat, at least along certain directions. The smaller the curvature of the manifold in the vicinity of the true signal, the weaker are the requirements on the accuracy of waveform fitting. A detailed analysis of the template-manifold curvature will have to be carried out in the future to understand how precise a best fit has to be such that the respective subtraction noise can be projected out. 

If one wants to extrapolate this result to higher observation times and more CBCs, then one needs to separate the detection and estimation problem from the projection problem. Certainly, it will be much more difficult to detect all signals of a realistic foreground, and for a more realistic analysis scheme, like the hierarchical search, one has to study the influence of confusion noise on the quality of the waveform estimates. The final answer depends on how close one will come to an optimal CBC analysis. We do not intent to make any predictions here concerning this point. However, assuming that all CBCs are detected and accurately analyzed, we claim that the projected subtraction-noise spectrum of a realistic foreground observed over 3 years will be negligible. First, we argued in our paper that confusion noise is no issue in the subtraction-noise projection, since the projection is based on the total Fisher matrix of all projected signals including mutual correlations into the model. Only if excluded, correlations between signals manifest themselves as confusion noise which may deteriorate the projection. Second, Fisher matrices which characterize 3 years of data will be much less ill-conditioned and for that reason much easier to handle numerically. That is the main reason why we consider our projection results very promising: we obtained good results choosing templates with strong parameter correlations. Third, and more fundamentally, the projected spectral power is supposed to decrease with $1/T$ which further reduces the projected spectrum by a factor 1000. 

Now, applying simple, numerically confirmed scaling laws of the correlation SNR, we extrapolate correlation results obtained for a few runs with observation times below $T=10^5\,$s to the full BBO mission lifetime of $T=10^8\,$s. In conclusion, provided that the subtraction noise can be removed, the extrapolated sensitivity of BBO to a stochastic background is $\Omega_{\min} \approx 2.5\cdot 10^{-17}$. However, one condition for this result is that the cosmological WD/WD foreground above 0.2\,Hz can be resolved and fitted like the foreground of NSs and BHs. Conversely, it may turn out that the lower boundary of the correlation band is too pessimistic which would entail a better sensitivity of BBO (up to a factor of 2). Ultimately, the problem of the WD barrier can be ameliorated making the BBO arms shorter by a small factor to increase the frequency of optimal sensitivity towards stochastic backgrounds. 

\section{Acknowledgments}
First of all, we want to thank Curt Cutler for many helpful discussions over the last year. Thanks to Bruce Allen and all members of his group who were always willing to discuss the BBO. Finally, J.~H.~expresses his gratitude to the Max-Planck society for funding his work in the exciting field of future-generation gravitational-wave detection.

\appendix

\section{Parameter derivatives of GW phase}
\label{secAppDer}
In this appendix, we elaborate on some details concerning the Fisher matrix of comparable-mass compact-binary inspirals within the pN formalism. The general framework is set in section \ref{secFisher}. 
 
The GW phase evolution depends on the three parameters $t_{\rm c},\,M,\,\mu$, and therefore the respective Fisher-matrix components depend on the derivative of the GW phase with respect to these parameters. We found that the result can be cast into a form which resembles the pN expansion Eq.~(\ref{eqPNphi}). In the following, these results will be presented. We start with the derivative of the phase with respect to the chirp time. It can be written as a sum
\beq
\partial_{t_c}^{\phantom{a}}\phi=-\frac{c^3}{4GM}\sum\limits_{k=0}^7p^{t_c}_k\tau^{-(3+k)/8}
\eeq
with expansion coefficients
\begin{widetext}
\beq
\begin{split}
p^{t_c}_0 & = 1\\
p^{t_c}_1 & = 0\\
p^{t_c}_2 & = \frac{743}{2688}+\frac{11}{32}\eta\\
p^{t_c}_3 & = -\frac{3}{10}\pi\\
p^{t_c}_4 & = \frac{1855099}{14450688}+\frac{56975}{258048}\eta+\frac{371}{2048}\eta^2\\
p^{t_c}_5 & = -\left(\frac{7729}{21504}+\frac{3}{256}\eta\right)\pi\\
p^{t_c}_6 & = -\frac{720817631400877}{288412611379200}+\frac{107}{280}C+\frac{53}{200}\pi^2-\frac{107}{2240}\log\left(\frac{\tau(t)}{256}\right)+\left( \frac{25302017977}{4161798144}-\frac{451}{2048}\pi^2\right)\eta\\
& \phantom{=}\;-\frac{30913}{1835008}\eta^2+\frac{235925}{1769472}\eta^3\\
p^{t_c}_7 & = \left(-\frac{188516689}{433520640}-\frac{28099}{57344}\eta+\frac{122659}{1290240}\eta^2\right)\pi
\end{split}
\eeq
\end{widetext}
All symbols are defined as in section \ref{secFisher}. A similar expression can be found for the derivatives with respect to the two mass parameters. The two expansions
\begin{eqnarray}
\partial_{\mu}^{\phantom{a}}\phi&=\frac{3}{4\eta\mu}\sum\limits_{k=0}^7p^{\mu}_k\tau^{(5-k)/8}\\
\partial_{M}^{\phantom{a}}\phi&=\frac{1}{2\eta M}\sum\limits_{k=0}^7p^{M}_k\tau^{(5-k)/8}
\end{eqnarray}
are determined by expansion coefficients
\begin{widetext}
\beq
\begin{split}
p^{\mu}_0 & = 1\\
p^{\mu}_1 & = 0\\
p^{\mu}_2 & = \frac{18575}{24192}-\frac{55}{96}\eta\\
p^{\mu}_3 & = -\frac{3}{2}\pi\\
p^{\mu}_4 & = \frac{9275495}{6193152}-\frac{284875}{774144}\eta-\frac{5565}{2048}\eta^2\\
p^{\mu}_5 & = \left(\frac{38645}{64512}+\frac{5}{256}\eta-\frac{38645}{64512}\log\left(\frac{\tau(t)}{\tau(0)}\right)\right)\pi\\
p^{\mu}_6 & = \frac{818786359710637}{19227507425280}-\frac{321}{56}C-\frac{159}{40}\pi^2+\frac{321}{448}\log\left(\frac{\tau(t)}{256}\right)+\left( -\frac{126510089885}{12485394432}+\frac{2255}{6144}\pi^2\right)\eta\\
& \phantom{=}\;-\frac{154565}{786432}\eta^2+\frac{5898125}{1769472}\eta^3\\
p^{\mu}_7 & = \left(\frac{942583445}{260112384}+\frac{140495}{172032}\eta+\frac{122659}{258048}\eta^2\right)\pi
\end{split}
\eeq
\beq
\begin{split}
p^{M}_0 & = 1\\
p^{M}_1 & = 0\\
p^{M}_2 & = -\frac{3715}{8064}+\frac{55}{32}\eta\\
p^{M}_3 & = \frac{3}{2}\pi\\
p^{M}_4 & = -\frac{9275495}{4816896}+\frac{284875}{258048}\eta+\frac{9275}{2048}\eta^2\\
p^{M}_5 & = \left(-\frac{38645}{21504}-\frac{15}{256}\eta+\frac{38645}{43008}\log\left(\frac{\tau(t)}{\tau(0)}\right)\right)\pi\\
p^{M}_6 & = -\frac{808989486879661}{11536504455168}+\frac{535}{56}C+\frac{53}{8}\pi^2-\frac{535}{448}\log\left(\frac{\tau(t)}{256}\right)+\left( \frac{126510089885}{4161798144}-\frac{2255}{2048}\pi^2\right)\eta\\
& \phantom{=}\;+\frac{463695}{1835008}\eta^2-\frac{8257375}{1769472}\eta^3\\
p^{M}_7 & = -\left(\frac{188516689}{28901376}+\frac{140495}{57344}\eta+\frac{122659}{258048}\eta^2\right)\pi
\end{split}
\eeq
\end{widetext}

\bibliography{references}

\end{document}